\documentclass[pra,aps,amsmath,reprint,10pt,twocolumn]{revtex4-2}

\usepackage{comment}
\usepackage{natbib}       
\usepackage{graphicx}     
\usepackage{xcolor}       
\usepackage{amsfonts}     
\usepackage{amssymb}      
\usepackage{amsthm}       
\usepackage{bbold}        
\usepackage{bbm}          
\usepackage{hyperref}     
\usepackage[normalem]{ulem}
\hypersetup{
    colorlinks=true,
    linkcolor=blue,
    citecolor=blue,
    filecolor=magenta,
    urlcolor=cyan,
}

\DeclareMathOperator{\Tr}{Tr}
\newcommand{\ket}[1]{\vert #1 \rangle}
\newcommand{\bra}[1]{\langle #1 \vert}
\newcommand{\braket}[2]{\langle #1 \vert #2 \rangle}

\begin{document}

\title{
\textcolor{black}{ Violations of the Leggett–Garg inequality in Hybrid Liouvillian Dynamics:\\ The Nonlinear Role of Detector Efficiency}}

\newcommand*{\affaddr}[1]{#1}
\newcommand*{\affmark}[1][*]{\textsuperscript{#1}}

\author{Sourav Paul\affmark[1]}
\email{sp20rs034@iiserkol.ac.in}
\author{Parveen Kumar \affmark[2]}
\email{parveenkumar622@gmail.com}
\author{Sourin Das \affmark[1]}
\email{sdas.du@gmail.com;sourin@iiserkol.ac.in}

\affiliation{
\affaddr{\affmark[1] Indian Institute of Science Education and Research Kolkata, Mohanpur, Nadia 741246, West Bengal, India}\\
\affaddr{\affmark[2] Department of Physics, Indian Institute of Technology Jammu, Jammu 181221, India}}

\begin{abstract}

Violations of the Leggett–Garg inequality (LGI) up to its algebraic bound under non-Hermitian dynamics are well established theoretically. Here, we demonstrate that such extreme violations are intrinsically fragile when realistic measurement processes are taken into account. We consider an open two-level system described by a time-local hybrid Liouvillian, with a continuous parameter \( q \in [0,1] \) representing detector efficiency, i.e., the fraction of quantum jump trajectories that are retained in the ensemble. This parameter interpolates between trace-preserving Lindblad dynamics (\(q=1\)) and non-Hermitian “no-jump” evolution (\(q=0\)). While \(K_3\) approaches its algebraic maximum of 3 in the null-efficiency limit, even an infinitesimal increase in detector efficiency induces a rapid, highly nonlinear suppression toward the classical bound. This logarithmic sensitivity reveals that maximal LGI violations are not robust physical features but rather singular limits of idealized measurement conditions. Our results have direct experimental implications: achieving algebraic LGI violations in systems undergoing continuous time evolution requires near-perfect suppression of detected quantum jumps (i.e., effective post-selection), placing stringent constraints on detector performance. In contrast to discrete protocols based on time-non-divisible dynamics, our framework shows that extreme violations arising within continuous, divisible quantum trajectory evolution constitute a fundamentally fragile regime.

\end{abstract}
\maketitle

\section{Introduction}
\label{I}

The boundary between classical macrorealism and quantum superposition remains a central and stringently tested question in modern physics~\cite{Leggett2002,Schlosshauer2005,Zurek2003,Kofler2007}. While microscopic systems readily exhibit superposition and entanglement, macroscopic objects appear to possess definite properties independent of observation. Leggett and Garg formalized this tension through inequalities designed to test the compatibility of quantum mechanics with the principles of \textit{macrorealism} (MR) and \textit{non-invasive measurability} (NIM)~\cite{LG1985,Leggett1988}. Violations of the  Leggett--Garg inequality (LGI) under loophole-free condition\cite{Kofler2013,Hensen2015}, thus provide an operational witness of non-classical temporal correlations, fundamentally analogous to Bell inequality violations for spatial entanglement~\cite{Bell1964,Brunner2014,Emary2014}.

Experimentally, LGIs are probed by sequentially measuring a dichotomic observable $\hat{Q} = \pm 1$ at times $t_1, t_2, t_3$, from which the temporal correlators $C_{ij} = \langle \hat{Q}(t_i)\hat{Q}(t_j)\rangle$ are constructed. These correlators define the standard LGI parameter $K_3 = C_{12} + C_{23} - C_{13}$, which classically satisfies the bound $-3 \le K_3 \le 1$~\cite{LG1985,Emary2014}. Violations of this classical limit have been successfully observed across diverse physical platforms, including superconducting circuits~\cite{Palacios2010,Groblacher2015}, nuclear magnetic resonance systems~\cite{Katiyar2013,Mouse2017}, and nitrogen-vacancy centers in diamond~\cite{Waldherr2011}. However, within the framework of standard quantum mechanics, the linearity of unitary evolution and projective measurements strictly bounds this violation to the temporal Tsirelson bound (TTB), also known as the L\"uders bound, $(K_3)_{\rm QM} \le 1.5$~\cite{Fritz2010,Budiyono2013,Maroney2014}. 

Surpassing the L\"uders bound necessitates modifying at least one of the foundational assumptions of standard Hermitian, i.e., trace-preserving dynamics. Historically, several theoretical approaches have been proposed to achieve this, including the utilization of non-Hermitian  and $\mathcal{PT}$-symmetric Hamiltonians~\cite{Varma2021,KumariPan1,KumariPan2}, weak and generalized measurement schemes~\cite{Dressel2014,Knee2016}, and post-selected dynamics~\cite{Jordan2015, VarmaYogesh}. Violation of L\"uders bound has also been discussed in the context of intrinsically non-divisible, discrete time evolution models~\cite{SpaulJPhysA, SpaulPRAletters}. A recent experimental progress in this direction was reported in ~\cite{MaheshUshaDevi2024}. They demonstrated that subjecting a system to a superposition of unitary operators yields macroscopic LGI violations well beyond the TTB, approaching the algebraic maximum of $K_3 = 3$. However, this map-based approach is intrinsically non-divisible; it treats time as discrete iterative steps rather than continuous evolution and therefore does not admit a time-local Markovian master equation description. This leaves open the question of the feasibility for extreme LGI violations (beyond TTB) within time-local dynamics under realistic conditions.

\begin{figure*}[t!]
\centering
\includegraphics[width=\textwidth]{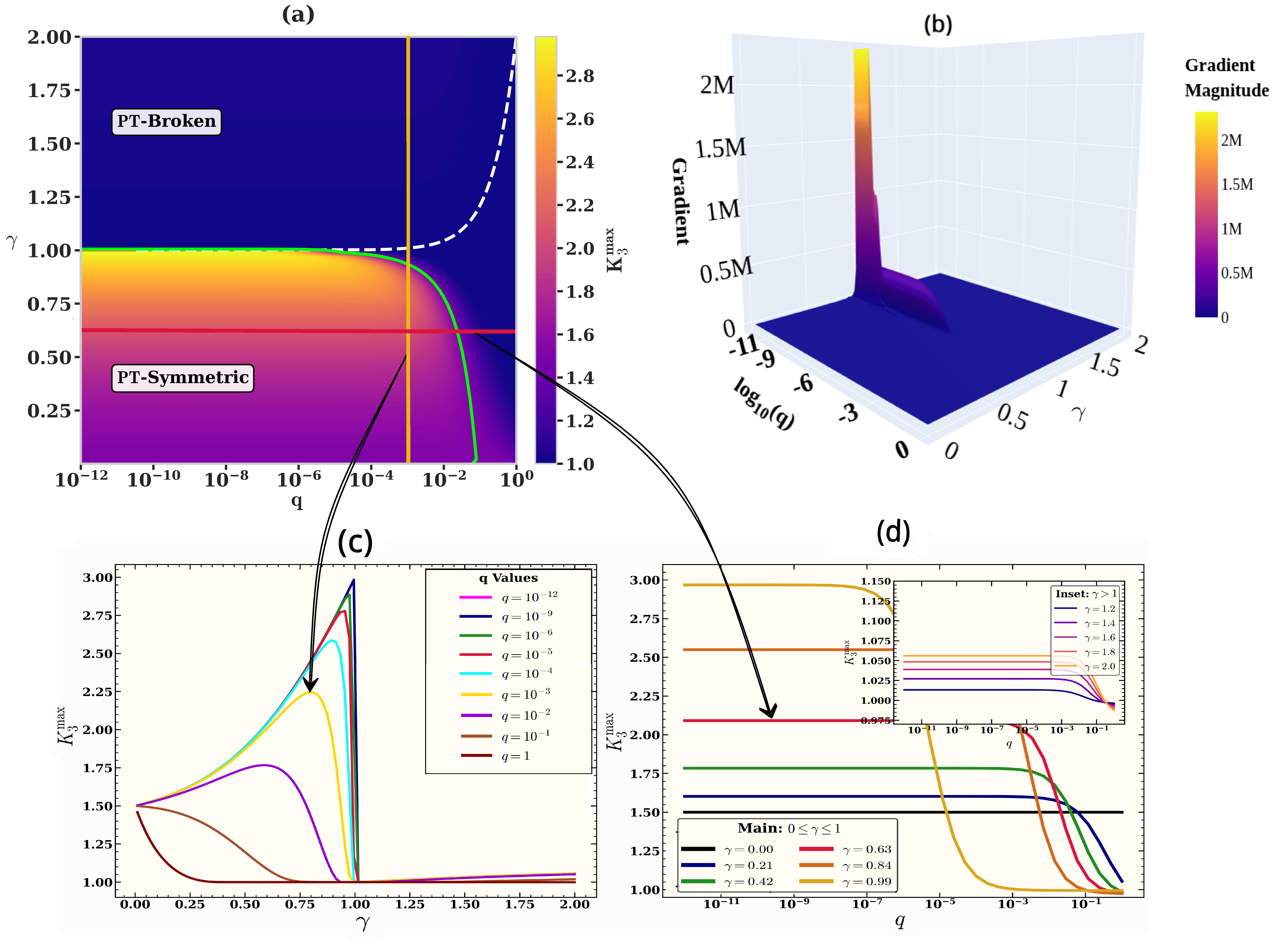}
\caption{ \textbf{Maximal violation of the LG parameter.} (a) The maximal value $K_3^{\max}$ is shown as a function of the detector efficiency parameter $q$ (displayed on a logarithmic scale) and the dissipation ratio $\gamma/J$. The colormap encodes the magnitude of $K_3^{\max}$, highlighting the parameter regions where the violation exceeds the L\"{u}ders bound. The green curve marks the boundary at $K_3 = 1.5$: the region below this line corresponds to super-quantum violation, while the region above corresponds to the quantum-allowed domain. The white dashed curve denotes the exceptional line of the hybrid Liouvillian [Eq.~(\ref{E01})], with further discussion provided in Appendix~\ref{app:liouvillian_spectrum}. (b) The corresponding gradient of $K_3^{\max}$ over the same parameter plane illustrates the descent structure of the landscape. \textbf{One-dimensional dependence of $K_3^{\text{max}}$ on $\gamma$ and $q$}: (c) The behavior of $K_3^{\text{max}}$ is shown as a function of the dissipation ratio $\gamma/J$ for fixed values of the detector efficiency parameter $q$. Each curve corresponds to a different choice of $q$, illustrating how detector efficiency parameter modulates the $\gamma$-dependence of the maximal LGI violation. (d) The variation of $K_3^{\text{max}}$ with respect to $q$ (displayed on a logarithmic scale) is plotted for fixed values of $\gamma/J$. Distinct curves represent different $\gamma$ values, demonstrating the sensitivity of the LGI violation to changes in the detector efficiency parameter across several orders of magnitude. In all panels, $J=1$, and $K_3$ is optimized over the measurement time interval $t$}  
\label{fig1Ch8}
\end{figure*}

Here we explore a complementary approach based on open quantum system under continuous monitoring to bring in the time-local framework~\cite{Minganti2019,Kumar2020,Kumar2021}. Starting from the standard Gorini--Kossakowski--Sudarshan--Lindblad (GKSL) master equation~\cite{Gorini1976,Lindblad1976,Breuer2002,Plenio1998}, we introduce a ``Hybrid Liouvillian'' framework. Central to this formalism is a crucial parameter $q \in [0,1]$, which naturally arises in quantum trajectory descriptions~\cite{Carmichael1993,Wiseman2009,Jacobs2014} and explicitly represents \textit{detector efficiency}. Physically, the $q$-parameter serves as a direct, continuous bridge between two distinct limits (Non-Hermitian and Liouvillian), acting as a physical knob that reflects the laboratory's inability to perfectly record all quantum jump events. 
We demonstrate that while the detector-efficiency parameter in the null limit yields $K_3$ violations approaching the algebraic maximum of 3, this convergence does not represent an optimal physical result. Instead, the central finding of this study is the extreme {fragility} of the limit as the efficiency parameter increases. Our analysis shows that these maximal violations are highly sensitive to fluctuations of detector efficiency, in fact they are logarithmically sensitive to realistic imperfections in quantum-jump detection, thereby posing severe challenges for experimental observation.

The structure of this paper is as follows: In Sec.~\ref{II}, we introduce our theoretical model, defining the system Hamiltonian and detailing the hybrid Liouvillian formalism for calculating two-time correlators in a non-trace-preserving context. In Sec.~\ref{III}, we present and discuss our numerical results, illustrating the rich, nonlinear behavior of $K_3$ as a function of the detector efficiency parameter $q$. Sec.~\ref{IV} is devoted towards the status violation of loophole-free conditions of macrorealism. Finally, in Sec.~\ref{V}, we summarize our findings and discussion.

\section{Hybrid Liouvillian and Detector Efficiency}
\label{II}

We consider a single qubit evolving under a continuous-time hybrid Liouvillian master equation:
\begin{equation}
\frac{d\rho}{dt} = -i [H, \rho] + 2\gamma \left( q\, L \rho L^{\dagger} - \frac{1}{2} \{ L^{\dagger} L, \rho \} \right).
\label{E01}
\end{equation}
Here, $\rho$ is the density matrix, $H = -(J/2)(\sin\theta\, \sigma_x + \cos\theta\, \sigma_z)$ governs the coherent tunneling dynamics (we set $\theta=\pi/2$ hereafter, so $H = -(J/2)\sigma_x$ and we put $J=1$ hereafter for all calculations), $L = \ket{\uparrow}\bra{\downarrow} = \sigma_+$ is the quantum jump operator representing environmental dissipation, and $\gamma$ is the dissipation rate. 

The crucial innovation in this model is the dimensionless parameter $q \in [0,1]$, which acts as a physical dial for detector efficiency during continuous monitoring. In a standard open system ($q=1$), the environment is an unobserved information sink; every quantum jump is lost, yielding trace-preserving Gorini-Kossakowski-Sudarshan-Lindblad (GKSL) evolution. Conversely, if the environment is continuously monitored with perfect efficiency, we can selectively filter the ensemble. For $q=0$, we post-select solely the ``null-measurement'' trajectories where no jumps occur. This removes the recycling term ($L \rho L^\dagger$), reducing the dynamics to a purely non-Hermitian effective Hamiltonian $H_{\text{eff}} = H - i\gamma L^{\dagger}L$. Intermediate values ($0 < q < 1$) model the realistic laboratory scenario: imperfect detector efficiency. Here, $q$ is the probability fraction that a quantum jump trajectory is retained (i.e., not post-selected out) of the ensemble. Because the evolution for $q \neq 1$ is non-trace-preserving, physical observables rely on the dynamically normalized state $\tilde{\rho}(t) = \rho(t) / \Tr[\rho(t)]$. This continuous renormalization couples the trace decay back into the Bloch vector components, rendering the state update equation highly nonlinear. This framework is formally derived from a microscopic system-detector interaction in the continuum limit, as detailed in Appendix~\ref{app:derivation1}.

\begin{figure*}[t!]
\includegraphics[width=0.99\textwidth]{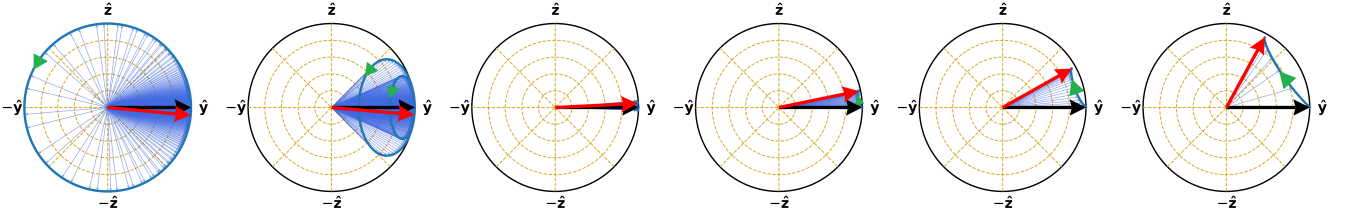}
\caption{\textbf{Bloch vector trajectories under hybrid Liouvillian dynamics.} 
Evolution of the quantum state projected onto the $\hat{y}$-$\hat{z}$ plane, initialized in the $\vert +\rangle_y$ state. 
System parameters are fixed at $J=1$, $\gamma = 0.9905$. 
Black and red markers indicate the initial ($t=0$) and asymptotic ($t \to \infty$) states, respectively. 
Blue markers trace the intermediate temporal evolution, while green arrows depict the instantaneous flow direction of the trajectory. 
Plots correspond to varying parameter values arranged from left to right: $q \in \{0, 10^{-5}, 10^{-3}, 10^{-2}, 10^{-1}, 1\}$. 
A distinct qualitative transition in the dynamical topology is observable as $q$ increases.}
\label{fig2Ch8}
\end{figure*}
\subsection{Calculation of Temporal Correlation and Leggett-Garg Parameter}
\label{IIA}

To construct temporal correlation and thereafter LGI pqarameter, we consider the dichotomic observable $Q = \sigma_y$, with the system initialized in the eigenstate $\rho(0) = \ket{+_y}\bra{+_y}$, where $\ket{+_y} = \frac{1}{\sqrt{2}}(\ket{\uparrow} + i\ket{\downarrow})$. A projective measurement of $Q$ at any time $t_k$ collapses the system onto the eigenstates $\ket{\pm_y}$ with the corresponding projectors $P_{\pm} = \ket{\pm_y}\bra{\pm_y}$. The system  evolves under the hybrid dynamical map $\mathcal{E}_t$ generated by Eq.~(\ref{E01}).

For the case where $q \neq 1$, the evolution is non-trace-preserving. Consequently, physical expectation values must be evaluated using the normalized state $\tilde{\rho}(t) = \mathcal{E}_t(\rho(0) / \Tr[\mathcal{E}_t(\rho(0)]$. To rigorously define the Leggett--Garg parameter $K_3(t)$, we evaluate three distinct two-time correlation functions using standard sequential-measurement protocols. The initial correlation $C_{01}(t)$ represents the correlation between the initial preparation at $t_0 = 0$ and the first measurement at $t_1=t$. The sequential correlation $C_{12}(t)$ accounts for the intermediate measurement at $t_1$ and the subsequent conditional evolution to $t_2=2t$. Finally, the unmeasured correlation $C_{02}(t)$ measures the correlation between $t_0=0$ and $t_2=2t$ in the absence of any intermediate measurement.

We construct the two-time correlation functions by decomposing them into joint probabilities of obtaining specific outcomes at different time intervals. In our quantum mechanical treatment, the joint probability $P(q_i, q_j)$ of obtaining outcome $q_i$ at time $t_1$ and $q_j$ at time $t_2$ is defined as the product of the probability of the first outcome, $p_{q_i}(t) = \Tr[P_{q_i} \tilde{\rho}(t)]$, and the conditional expectation of the second. This requires evaluating the normalized post-measurement state, 
\begin{equation}
\tilde{\rho}_{q_i}(t) = \frac{\mathcal{E}_t(P_{q_i})}{\Tr[\mathcal{E}_t(P_{q_i})]},
\end{equation}
which represents the system after the first measurement, having evolved for an additional duration $t$.

Using this probabilistic framework, the three required correlation functions $C_{ij}$ emerge naturally as expectation values of the outcome products. The initial correlation between the preparation and the first measurement at $t_1=t$ evaluates directly to $C_{01}(t) = \Tr[\sigma_y \tilde{\rho}(t)]$. For the sequential correlation, which accounts for the intermediate measurement at $t_1$ and the subsequent evolution to $t_2=2t$, we take the conditional average over both possible intermediate outcomes, yielding
\begin{equation}
C_{12}(t) = \Tr[\sigma_y \tilde{\rho}_{+}(t)] p_{+}(t) - \Tr[\sigma_y \tilde{\rho}_{-}(t)] p_{-}(t).
\end{equation}
Finally, the unmeasured correlation between the initial state at $t_0 = 0$ and the final time $t_2=2t$, assuming no intermediate measurement occurred, simplifies to $C_{02}(t) = \Tr[\sigma_y \tilde{\rho}(2t)]$.

The Leggett--Garg parameter is then defined as the linear combination of these correlations,
\begin{equation}
K_3(t) = C_{01}(t) + C_{12}(t) - C_{02}(t).
\end{equation}
For any macrorealistic theory, the condition $K_3 \leq 1$ must be satisfied. Violations of this bound ($K_3 > 1$) indicate the presence of quantum coherence and non-invasive measurability issues, which are significantly influenced by the detector efficiency parameter $q$. Fig.~\ref{fig1Ch8} illustrates the numerical result of the optimal value of $K_3$ in a comprehensive parameter scan.

\subsection{Bloch-Vector Representation}
\label{IIB}

The hybrid Liouvillian dynamics generated by Eq.~(\ref{E01}) admits a transparent representation in terms of the Bloch vector. Following the formalism for non-Hermitian PT-symmetric dynamics~\cite{Varma2021}, we decompose the density matrix as
\begin{equation}
\rho(t) = \frac{R(t)}{2}\,\mathbb{I} + \frac{1}{2}\,\vec{S}(t)\cdot\vec{\sigma},
\label{eq:bloch_decomp}
\end{equation}
where $R(t) = \mathrm{Tr}[\rho(t)]$ denotes the time-dependent trace and $\vec{S}(t) = \mathrm{Tr}[\rho(t)\vec{\sigma}]$ is the unnormalized Bloch vector. The physical, normalized state is described by $\vec{s}(t)=\vec{S}(t)/R(t)$ whenever $R(t) \neq 0$.

Substituting this decomposition into the master equation yields a set of coupled differential equations governing the evolution of the trace and the Bloch vector components. A key feature of these equations is that the trace is conserved ($\dot{R}=0$) only in the Lindbladian limit $q=1$. For any $q \neq 1$, the trace dynamics couples back into the spin evolution. Because the physical state depends on the ratio of $\vec{S}(t)$ to $R(t)$, this coupling renders the effective dynamics of the normalized vector $\vec{s}(t)$ highly nonlinear. We present the evolution of the normalized Bloch vector in Fig.~\ref{fig2Ch8}, with further mathematical details provided in Appendix~\ref{app:bloch_special} which also includes analytic expressions (with some approximations) of these Bloch spin vector components.

\section{Nonlinear Amplification of Temporal Correlations}
\label{III}

We now examine the numerical results for the Leggett--Garg parameter, defined via the three-time measurement protocol at $t_0=0$, $t_1=t$, and $t_2=2t$:
\begin{equation}
    K_3 = C_{01} + C_{12} - C_{02}.
\end{equation}
The parameter $K_3$ is numerically evaluated via joint probability conditions and optimized over the measurement interval $t$, the dissipation ratio $\gamma/J$, and the detector efficiency $q$. Our central finding is the extreme, highly nonlinear sensitivity of $K_3$ to $q$, as summarized in Fig.~\ref{fig1Ch8}.

\textit{\underline{Limiting Dynamics:}} The behavior of the LGI violation is anchored by two distinct physical regimes. In the \textbf{Lindblad Limit ($q = 1$)}, standard dissipation monotonically suppresses quantum coherence. As shown in Fig.~\ref{fig2Ch8}, state trajectories collapse toward the mixed-state origin of the Bloch sphere. In this regime, $K_3$ strictly obeys the L\"uders bound ($K_3 \le 1.5$) and trends toward the classical limit ($K_3 \to 1$) as $\gamma$ increases.
\begin{figure*}[t!]
\centering
\includegraphics[width=0.32\textwidth]{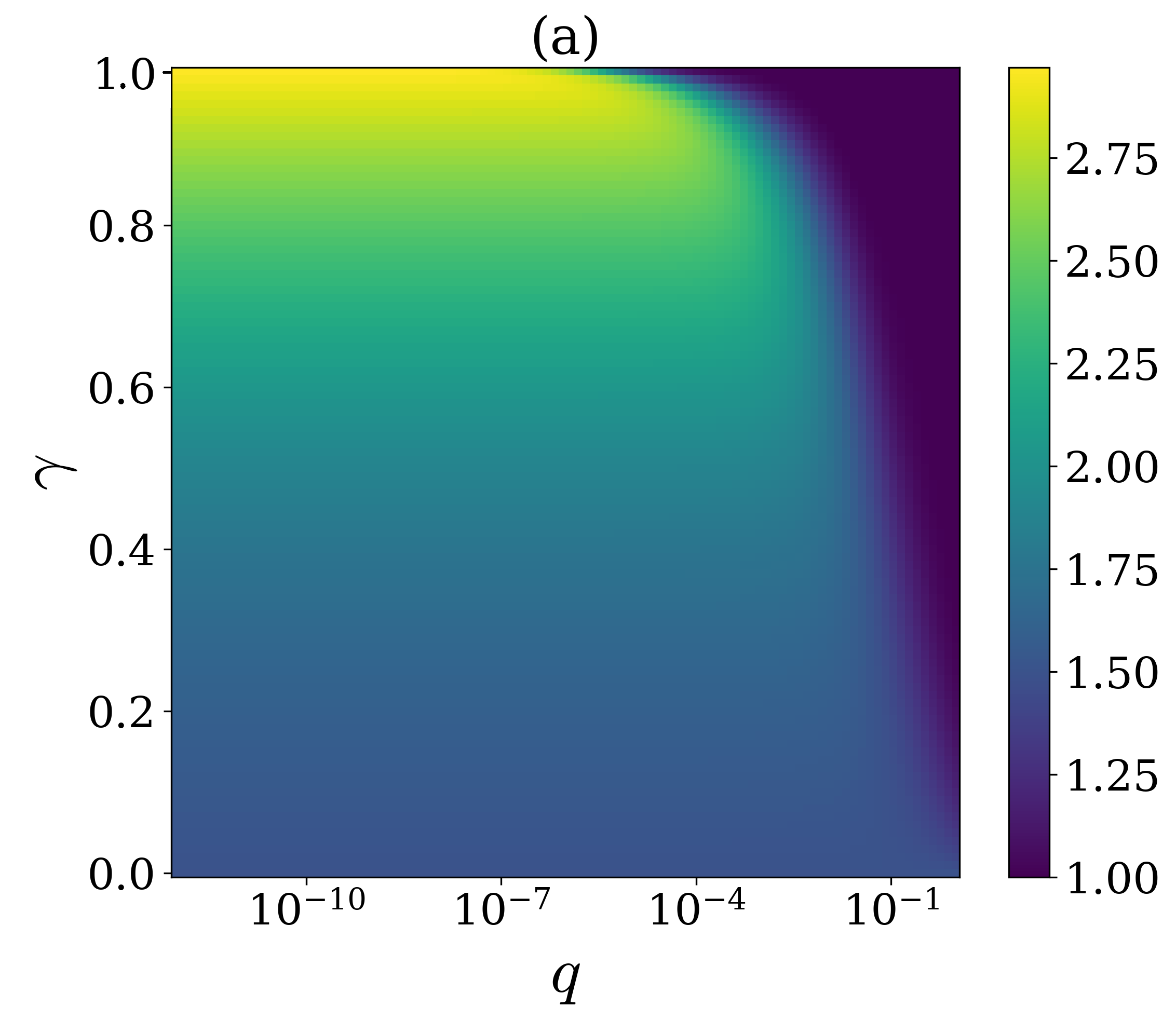}
\includegraphics[width=0.32\textwidth]{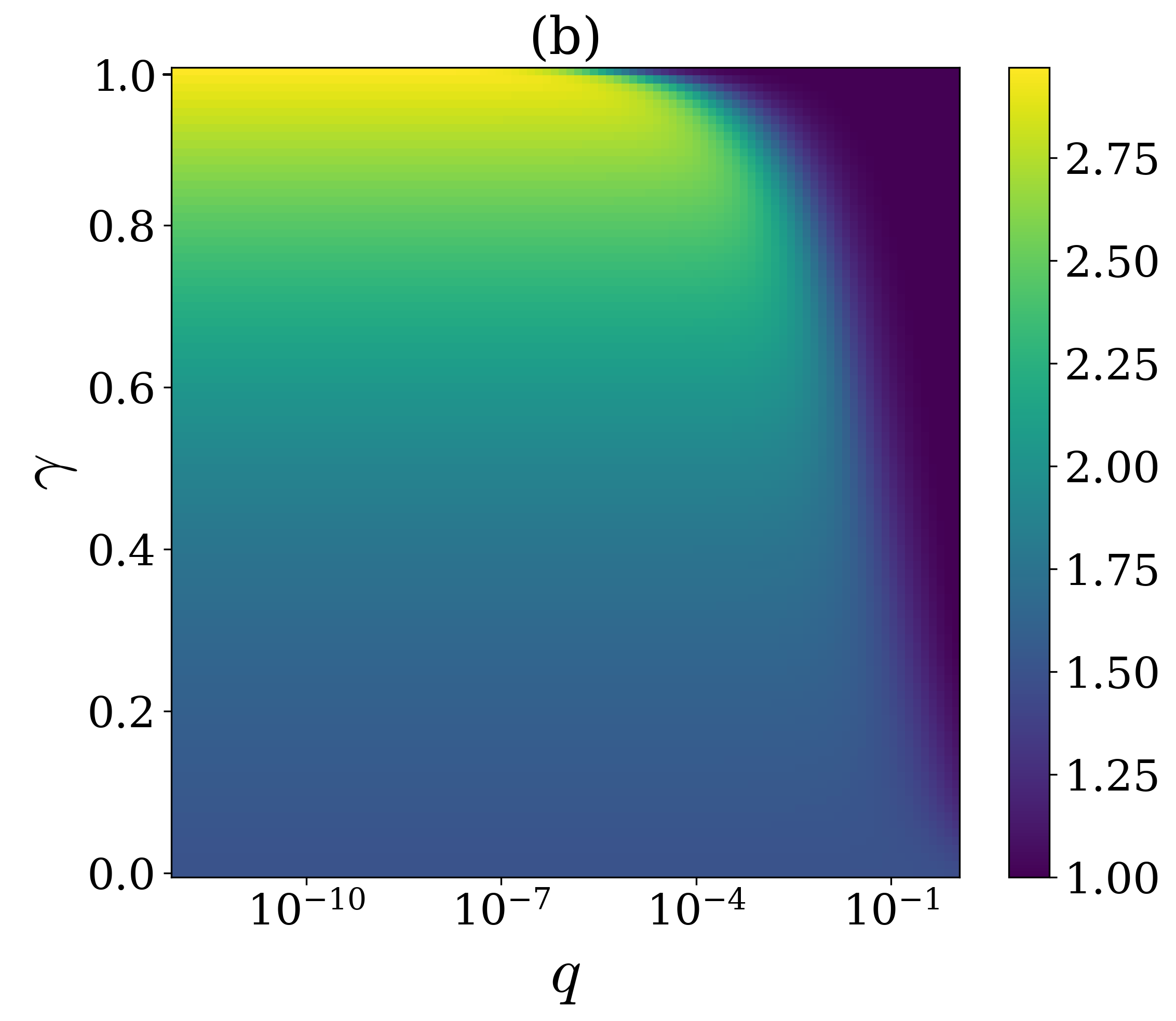}
\includegraphics[width=0.32\textwidth]{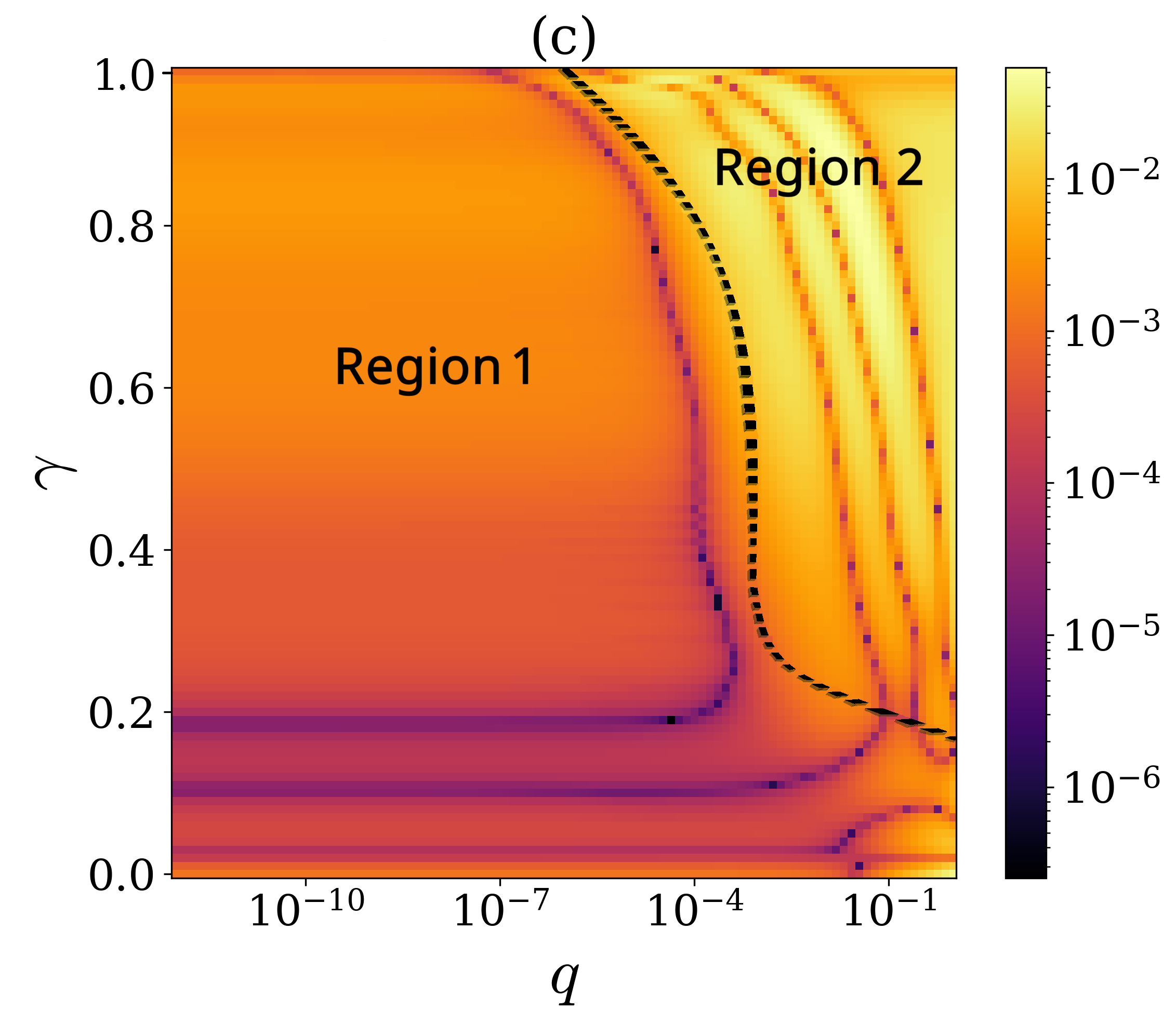}
\caption{ \textbf{Fitted curve for $K_3^{\text{max}}$.} Plot.(a) \ displays $K_3^{\text{max}}$ as a function of the parameters $\gamma$ and $q$. The interval $1 \leq \gamma \leq 2$ is not included, as this range exhibits no notable structure relevant for the fitting. Plot.(b) \ illustrates the resulting fit, where $K_3^{\text{max}}(q,\gamma)$ is modeled as a polynomial function of $\gamma$ (up to twentieth order) combined with a hyperbolic-tangent dependence on $\log q$ . This composite form captures the rapid variation in $\log q$ and the  nonlinear dependence on $\gamma$.  Plot.(c) \ shows the accuracy of the fit | the absolute value difference of plot.(a) and plot.(b) is captured here. The black bold dotted line divides the plot into two regions | on an overall scenario, Region 1 shows better accuracy of fit with the absolute value difference being below $10^{-2}$, whereas Region 2 shows an accuracy of absolute difference between $10^{-1}$ and $10^{-2}$ . For plots (a) and (b), the plotlegends shows the numerical value of the $K_3^{\text{max}}$ function and for Plot (c) it shows the absolute value difference. (For fitting expressions, see Appendix~\ref{app:fit_parameter}) .}
\label{fig3Ch8}
\end{figure*}

Conversely, in the \textbf{Non-Hermitian Limit ($q \ll 1$)}, conditioning the ensemble on jump-free trajectories robustly stabilizes coherent evolution despite environmental coupling. The highly nonlinear renormalized $\tilde{\rho}(t)$ forces the Bloch vector to undergo accelerated, geometry-altering excursions (see Fig.~\ref{fig2Ch8}). In narrow parametric windows where the Liouvillian eigenvalues coalesce near the \textit{exceptional points} ($\gamma \sim J=1$), this accelerated angular motion of the bloch vector allows the sequential correlation functions to combine constructively. To be more precise, $C_{01}$ and $C_{12}$ behave in the same quanlitative manner having same slight damped  ocsillatory behavior with equal period whereas $C_{02}$ behaves with same qualitative manner but with a time period half to them. This drives $K_3$ far beyond the L\"uders bound, asymptotically approaching the algebraic maximum $K_3 \to 3$.

\textit{\underline{Intermediate Regime:}} The transition from extreme violation to classical bounds is profoundly nonlinear. One-dimensional parametric cuts (Fig.~\ref{fig1Ch8} (c-d)) reveal that for fixed $\gamma$, increasing detector inefficiency (tuning $q$ away from 0) triggers a sharp, logarithmic-scale collapse of $K_3$. We find that the global optimization landscape of the maximal LGI violation is universally characterized by a hyperbolic-tangent profile in logarithmic space:
\begin{equation}
    K_3^{\max}(\gamma,q) = A(\gamma) \tanh\big[B(\gamma)\log q + C(\gamma)\big] + D(\gamma).
\end{equation}
where $A, B, C,$ and $D$ are continuous polynomial functions of $\gamma$ (Fig.~\ref{fig3Ch8}). This profound non-linearity dictates that extreme non-classicality is exceptionally fragile; even minimal fractions of unrecorded quantum jumps (e.g., $q \sim 10^{-4}$) drastically suppress the constructive interference required for algebraic saturation.
\par The origin of this $K_3$ parameter amplification is inextricably linked to the topology of the underlying hybrid Liouvillian dynamics. Specifically, the system's proximity to the \textit{exceptional boundary line}, defined by $4(\gamma^2 - J^2)^3 = 27q^2\gamma^2 J^4$, separates regimes of overdamped decay from accelerated pseudo-coherent oscillations as dictated by each correlation functions. As $q$ is reduced, the bloch trajectories transition from damped motion to strongly accelerated evolution near this boundary. These results establish that imperfect post-selection within hybrid Liouvillian dynamics enables a dramatic amplification of temporal quantum correlations, allowing $K_3$ to approach its absolute upper bound through the geometric manipulation of the Bloch vector.



\section{NSIT and AoT Conditions}
\label{IV}

In this section, we critically re-evaluate the interpretation of $K_3 > 1$ as a definitive signature of non-classical dynamics. While recent literature demonstrates significant efforts to close both the noninvasive measurability \cite{AKPan, Majidy} and clumsiness \cite{Huffman} loopholes, our analysis focuses exclusively on the statistical formulations of the No-Signaling in Time (NSIT) and Arrow of Time (AoT) conditions. It is crucial to note that the simultaneous satisfaction of both the NSIT and AoT conditions is necessary to guarantee the existence of a global joint probability distribution \cite{Clemente}, a prerequisite that inherently implies macroscopic realism. 

\begin{figure*}[t!]
\centering
\includegraphics[width=0.98\textwidth]{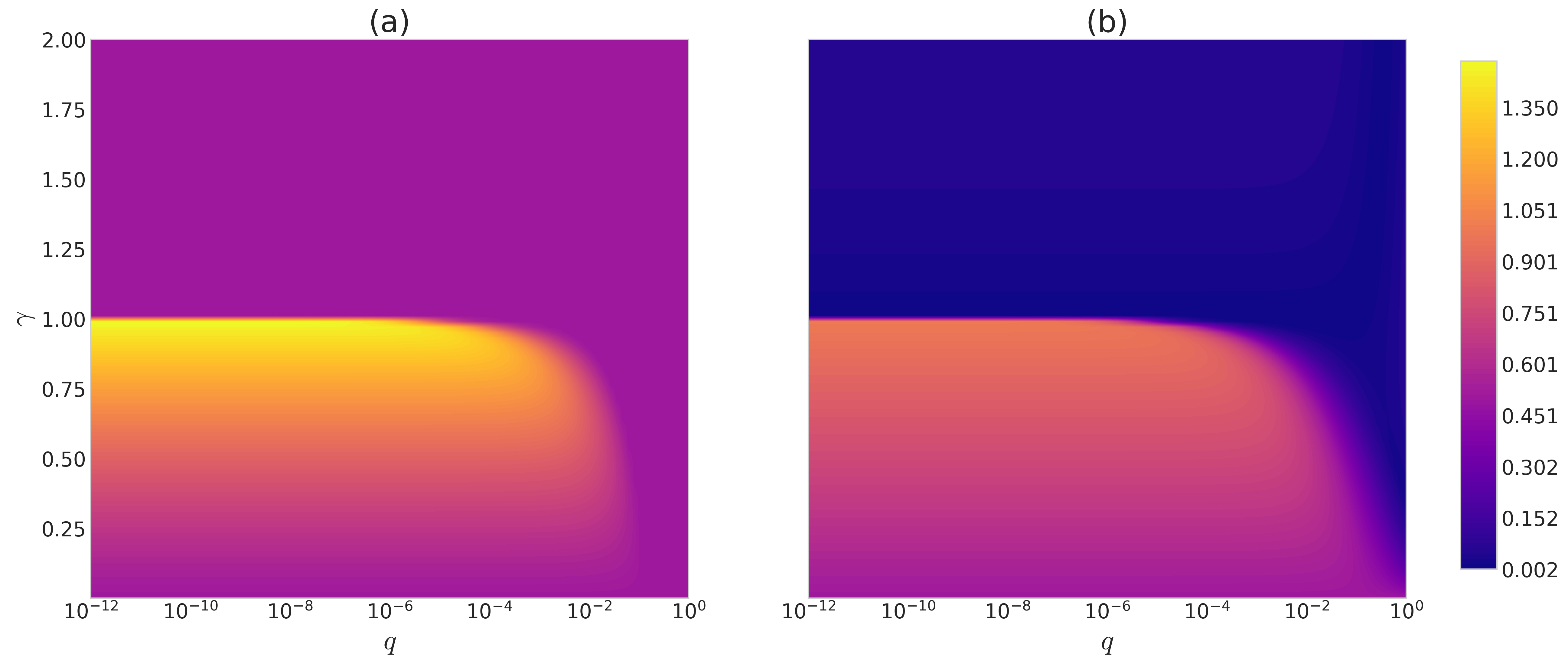}
\caption{\textbf{Violation of two and three-time NSIT conditions:} (a) Plot of a specific component of $\Delta_{0(1)2}$ as a function of the parameters $\gamma$ and $q$, corresponding to the dichotomic measurements $q_0 = +1$ at time $t_0 = 0$ and $q_2 = +1$ at time $t_2 = 2t$. (b) Plot illustrates a specific component of $\Delta_{(i)j}$ for $i=1$ and $j=2$, given the dichotomic measurement $q_2 = +1$ at time $t_2 = 2t$. Further details are provided in Appendix~\ref{app:NSIT_AoT}.}
\label{fig4Ch8}
\end{figure*}
Under standard unitary dynamics, it is well established that NSIT conditions are typically violated, whereas AoT conditions are preserved. Building upon this foundational understanding, we clarify the implications of hybrid Liouvillian dynamics for macroscopic realism by analyzing the behavior of these conditions within our model. For dichotomic outcomes $q_i=\pm1$ measured at successive times $t_0<t_1<t_2$, macroscopic realism is strictly equivalent to the existence of a well-defined global joint probability distribution $P(q_0,q_1,q_2)$. Consequently, this requires the simultaneous validity of all applicable NSIT and AoT constraints.

An explicit evaluation of the multitime probabilities generated by the hybrid Liouvillian reveals that all two- and three-time AoT conditions are identically satisfied. Specifically, the relations
\begin{align}
 \text{AoT}_{i(j)}: P(q_i) &= \sum_{q_j=\pm1} P(q_i,q_j), \qquad i<j, \\
 \text{AoT}_{01(2)}: P(q_0,q_1) &= \sum_{q_2=\pm1} P(q_0,q_1,q_2),
\end{align}
hold true for arbitrary parameter values. This universal satisfaction reflects the underlying causal structure of the dynamical map. 

In stark contrast, the NSIT conditions are generically violated. Within the hybrid Liouvillian framework governing the qubit system's evolution, the two-time NSIT relations,
\begin{equation}
\text{NSIT}_{(i)j}: P(q_j)=\sum_{q_i=\pm1}P(q_i,q_j), \qquad i<j,
\end{equation}
as well as the three-time NSIT relations,
\begin{align}
\text{NSIT}_{0(1)2}: P(q_0,q_2) &= \sum_{q_1=\pm1} P(q_0,q_1,q_2), \\
\text{NSIT}_{(0)12}: P(q_1,q_2) &= \sum_{q_0=\pm1} P(q_0,q_1,q_2),
\end{align}
exhibit broad violations across the general parameter space (see Appendix \ref{app:NSIT_AoT} for further details). Interestingly, for our specific model, $\text{NSIT}_{(0)12}$ is always satisfied. We therefore define the quantities $\Delta_{(i)j} = \vert P(q_j) - \sum_{q_i=\pm1}P(q_i,q_j) \vert$ and $\Delta_{0(1)2} = \vert P(q_0,q_2) - \sum_{q_1=\pm1} P(q_0,q_1,q_2)\vert$ to explicitly quantify the violation of the remaining NSIT conditions, as illustrated in Fig.~\ref{fig4Ch8}. 

The systematic failure of the NSIT conditions---despite the universal satisfaction of AoT constraints---across both the intrinsic physical dynamics (the full Lindbladian without post-selection) and the post-selected ensemble (hybrid Liouvillian) dynamics ultimately demonstrates that our hybrid Liouvillian framework is intrinsically incompatible with macroscopic realism.

\section{Discussion and Conclusion}
\label{V}

The hybrid Liouvillian framework establishes an experimentally pragmatic route for engineering extreme quantum temporal correlations in dissipative environments. By optimizing the Leggett--Garg parameter $K_3$ over the measurement interval $t$, dissipation strength $\gamma$, and detector efficiency $q$, we have demonstrated that imperfect post-selection serves as a potent, albeit delicate, resource for violating the Leggett--Garg inequality (LGI)~\cite{LG1985}.

Our results highlight two pivotal conclusions regarding the quantum-to-classical transition. First, we establish that the ability to push temporal correlations toward their algebraic limit $K_3 \to 3$ depends nonlinearly on the detector efficiency of environmental monitoring. Because $K_3$ exhibits a sharp, approximately logarithmic dependence on $q$, any departure from perfect post-selection rapidly introduces trace-preserving noise that penalizes non-classical amplification. This sensitivity is captured by our universal scaling form, where $K_3^{\max}(q, \gamma)$ follows a $\tanh(\log q)$ profile, implying that experimentalists must ensure extraordinarily high-fidelity detection to surpass the L\"uders bound of $1.5$~\cite{Luders1951, Budroni2014}.

Second, our findings provide a vital counterpoint to recent map-level superposition protocols. While Chatterjee et al.~\cite{MaheshUshaDevi2024} achieved violations of the temporal Tsirelson bound via a coherent linear combination of unitary propagators $A(t) = c_1 U_1(t) + c_2 U_2(t)$, such constructions result in inherently non-divisible maps. In contrast, our model confirms that discrete superpositions are not strictly necessary to exceed these bounds. Here, the requisite dynamical nonlinearity is natively generated by the back-action of continuous null-measurements within a fully divisible, time-local master equation. The extreme violations observed near the Liouvillian exceptional points arise from accelerated, geometry-altering excursions on the Bloch sphere~\cite{Rotter2009, Daley2014}, a feature unique to this hybrid dissipative-conditioned evolution.

In summary, by translating the abstract concept of non-Hermitian trajectory filtering into the operational metric of detector efficiency, this work shifts the pursuit of extreme macrorealism violations from discrete map engineering to the continuous quantum control of open systems. Lastly, we briefly discussed the status of violation of No-Signaling in Time (NSIT) and AoT (arrow of time) conditions required to address the loophole-free conditions providing a comprehensive framework for tests of macroscopic realism for non-trace-preserving Hybrid Liouvillian dynamics, in state-of-the-art superconducting and photonic architectures.  

\section{Acknowledgement} 
S.P. offers his gratitude to the Council of Scientific and Industrial Research (CSIR), Govt. of India for financial support. S.D. would like to acknowledge the financial support from Anusandhan National Research Foundation (ANRF) under the MATRICS scheme (Grant No. [ANRF/ARGM/2025/002511/TS)]);  Ministry of Education, Government of India under the SPARC program (Project Code: [SPARC/2025-2026/P4086]) and National Quantum Mission under Quantum Algorithms Technical Group (TPN No.: 136428). PK acknowledges support from the IIT Jammu Initiation Grant No. SGT-100106.

\appendix
\texttt
\clearpage 
\onecolumngrid 
\begin{center}
    \textbf{\large APPENDIX}
\end{center}
\onecolumngrid 

\section{Microscopic Derivation of the Hybrid Dynamics}
\label{app:derivation1}

In this appendix, we provide the microscopic foundation for the hybrid Liouvillian model utilized in the main text. We proceed by defining the system-detector interaction and deriving the continuous-time master equation by explicitly tracing out the detector degrees of freedom and incorporating imperfect post-selection.

\subsection{System-Detector Hamiltonian}

We consider a system coupled to a detector, modeled as a two-level quantum object. The system Hamiltonian is given by $H_s = \omega \hat{n} \cdot \boldsymbol{\sigma}$, where $\omega$ is the Zeeman energy, and $\hat{n}$ is a unit vector parametrized by spherical coordinates $(\theta, \phi)$. The detector is prepared in the initial state $\rho_d^0 = \frac{1}{2}(\mathbb{I} + \hat{m} \cdot \boldsymbol{\sigma})$, where $\hat{m}$ defines the initialization direction.

The joint system-detector state, initially $\rho(t) = \rho_s(t) \otimes \rho_d^0$, evolves under the interaction Hamiltonian
\begin{equation}
H_{s-d} = \sqrt{2}J [\sigma_x^s \otimes \sigma_x^d + \sigma_y^s \otimes \sigma_y^d ],
\end{equation}
where $J$ is the coupling strength. This anisotropic interaction couples the system's spin direction orthogonal to the detector's initialization. Following the interaction, the detector is measured projectively and reset.

\subsection{Kraus Representation and Imperfect Post-Selection}

The reduced dynamics of the system over an infinitesimal time step $\delta t$ is obtained by tracing out the detector degrees of freedom. Let $U(\delta t) = \exp[-i(H_s \otimes \mathbb{I} + H_{s-d})\delta t]$ be the joint unitary evolution operator. The updated system state is
\begin{equation}
\tilde{\rho}_s(t+\delta t) = \Tr_d \left[ U(\delta t) (\rho_s(t) \otimes \rho_d^0) U^\dagger(\delta t) \right].
\end{equation}
Evaluating the partial trace in the eigenbasis $\{|k\rangle_d\}$ of the detector allows us to express the evolution in terms of Kraus operators $M_k = {}_d\langle k | U(\delta t) | \psi_{init} \rangle_d$, such that $\tilde{\rho}_s(t+\delta t) = \sum_k M_k \rho_s(t) M_k^\dagger$.
For the specific interaction chosen, this procedure yields two dominant operators representing the "no-jump" ($M_0$) and "jump" ($M_1$) scenarios:
\begin{align}
M_0 &= \mathbb{I} - i H_{\text{eff}} \delta t, \\
M_1 &= \sqrt{2\gamma \delta t} \, L,
\end{align}
where $H_{\text{eff}} = H - i\gamma L^\dagger L$ and $L = \sigma_+$, with the effective rate $\gamma$ defined by the continuum scaling of the coupling $J$.

In a standard trace-preserving map ($q=1$), both outcomes are kept. To model imperfect post-selection, we introduce the efficiency parameter $q \in [0,1]$. This parameter scales the contribution of the jump trajectory, corresponding to the fraction of jump trajectories that are retained in the ensemble. The modified Kraus map for the unnormalized density matrix becomes:
\begin{equation}
\tilde{\rho}_{\text{S}}(t+\delta t) = M_0 \tilde{\rho}_{\text{S}}(t) M_0^{\dagger} + q M_1 \tilde{\rho}_{\text{S}}(t) M_1^{\dagger}.
\end{equation}
Substituting the explicit forms of $M_0$ and $M_1$:
\begin{align}
\tilde{\rho}_{\text{S}}(t+\delta t) &\approx (\mathbb{I} - i H_{\text{eff}} \delta t) \rho(t) (\mathbb{I} + i H_{\text{eff}}^\dagger \delta t) \nonumber \\
&\quad + 2\gamma q \delta t L \rho(t) L^\dagger.
\end{align}
Expanding to first order in $\delta t$, we obtain:
\begin{align}
\frac{\tilde{\rho}_{\text{S}}(t+\delta t) - \rho(t)}{\delta t} &\approx -i[H, \rho(t)] - \gamma \{L^\dagger L, \rho(t)\} \nonumber \\
&\quad + 2\gamma q L \rho(t) L^\dagger.
\end{align}
Taking the limit $\delta t \to 0$, we arrive at the hybrid Liouvillian master equation presented in Eq.~(\ref{E01}). This derivation explicitly demonstrates that the term $2\gamma q L \rho L^\dagger$ arises from the partial retention of the jump trajectories, while the anti-commutator term remains unaffected, preserving the coherent non-Hermitian decay characteristic of the no-jump evolution.

\section{Bloch Dynamics at $\theta = \pi/2$}
\label{app:bloch_special}

In this appendix, we provide the explicit equations of motion for the specific parameter regime $\theta = \pi/2$ used in the main text. The system Hamiltonian is $H = -(J/2)\sigma_x$, and the equation of motion of the Bloch vector in Sec.~\ref{IIB} takes following form. The evolution of the $S_x$ component decouples from the system:
\begin{equation}
\frac{d S_x}{dt} = -\gamma S_x.
\end{equation}
Given the initial state $\rho(0) = \ket{+_y}\bra{+_y}$, we have $S_x(0)=0$, which implies $S_x(t)=0$ for all $t$. The dynamics are therefore strictly confined to the $y$-$z$ plane.

The remaining coupled dynamics for the trace $R$ and the components $S_y, S_z$ can be written in compact matrix form as:
\begin{equation}
\frac{d}{dt}
\begin{pmatrix} R \\ S_y \\ S_z \end{pmatrix}
=
\begin{pmatrix}
-\gamma(1-q) & 0 & \gamma(1-q) \\
0 & -\gamma & J \\
\gamma(1+q) & -J & -\gamma(1+q)
\end{pmatrix}
\begin{pmatrix} R \\ S_y \\ S_z \end{pmatrix}
\label{EB2}
\end{equation}
This linear system describes the interplay between the coherent driving $J$ and the measurement-induced decay and recycling rates parameterized by $\gamma$ and $q$.
\subsection{Approx. Analytical Solution for the Coupled Spin System}
We rewrite the eq.(\ref{EB2}) with an approximation that we assume $ \gamma q S_z\simeq 0$ for the all equations and get the following form. {\it The numerical result for these equations (\ref{eq:bloch_R}) matches with the eqs.\ref{EB2}.}
\begin{align}
\frac{d R}{dt} &= \gamma S_z-\gamma(1-q)  R, \label{eq:bloch_R} \\
\frac{d S_y}{dt} &= J S_z  - \gamma S_y, \\
\frac{d S_z}{dt} &= -J S_y  - \gamma S_z + \gamma(1+q)R. 
\end{align}
Below we present the analytical solution for the state vector $\mathbf{v}(t) = [R(t), S_y(t), S_z(t)]^T$ based on the coupled differential equations with $J=1$ and initial conditions:

\begin{equation}
R(0) = 1, \quad S_y(0) = 1, \quad  S_z(0) = 0
\end{equation}

The system of equations is solved using the matrix method for the state vector $\mathbf{v}(t) = [R(t), S_y(t), S_z(t)]^T$. Given the initial conditions $\mathbf{v}(0) = [1, 1, 0]^T$.

\subsection*{1.1 The Characteristic Equation}

The eigenvalues $\lambda$ are found by solving the cubic equation derived from $\det(A - \lambda I) = 0$. Let $x = \lambda + \gamma$. The values of $x$ are the roots of:

\begin{equation}
x^3 + ax^2 + bx + c = 0
\end{equation}

where the coefficients are defined by the physical parameters:

\begin{align*}
a &= -\gamma q \\
b &= J^2 - \gamma^2(1+q) \\
c &= -\gamma q J^2
\end{align*}

\subsection*{1.2 Eigenvalues via Cardano's Method}

Defining the depressed cubic parameters:

\begin{equation}
P = \frac{3b - a^2}{3}, \quad Q = \frac{2a^3 - 9ab + 27c}{27}
\end{equation}

The intermediate terms for the exact roots are:

\begin{equation}
u = \sqrt[3]{-\frac{Q}{2} + \sqrt{\frac{Q^2}{4} + \frac{P^3}{27}}}, \quad v = \sqrt[3]{-\frac{Q}{2} - \sqrt{\frac{Q^2}{4} + \frac{P^3}{27}}}
\end{equation}

The three roots $x_j$ (and thus $\lambda_j = x_j - \gamma$) are:

\begin{align*}
x_1 &= u + v - \frac{a}{3} \\
x_2 &= \omega u + \omega^2 v - \frac{a}{3} \\
x_3 &= \omega^2 u + \omega v - \frac{a}{3}
\end{align*}

where $\omega = e^{i 2\pi / 3} = -\frac{1}{2} + i\frac{\sqrt{3}}{2}$.

\subsection*{1.3 Time-Domain Solutions}

Using the eigenvectors $\mathbf{u}_j = [\gamma x_j, J(x_j - \gamma q), x_j(x_j - \gamma q)]^T$, the exact solutions are:

\begin{align}
R^{+}(t) &= \gamma e^{ - \gamma t}\sum_{j=1}^3 c_j^{+} x_j e^{x_j t} \\
S_y^{+}(t) &= J e^{ - \gamma t}\sum_{j=1}^3 c_j^{+} (x_j - \gamma q) e^{x_j t} \\
S_z^{+}(t) &= e^{ - \gamma t}\sum_{j=1}^3 c_j^{+} x_j (x_j - \gamma q) e^{qx_j t}
\end{align}

The coefficients $c_j$ are determined by the initial conditions:

\begin{equation}
c_j^{+} = \frac{x_k x_l - \gamma q(x_k + x_l) + \gamma^2 q + \frac{\gamma}{J}x_k x_l}{\gamma(x_j - x_k)(x_j - x_l)}
\end{equation}

for $(j, k, l)$ as cyclic permutations of $(1, 2, 3)$.

\begin{equation} s_y^{+}(t) = \frac{J \sum_{j=1}^3 c_j^{+} (x_j - \gamma q) e^{x_j t}}{\gamma \sum_{j=1}^3 c_j^{+} x_j e^{x_j t}} \end{equation}

\begin{equation} 
s_z^{+}(t) = \frac{\sum_{j=1}^3 c_j^{+} x_j (x_j - \gamma q) e^{x_j t}}{\gamma \sum_{j=1}^3 c_j^{+} x_j e^{x_j t}} 
\end{equation}


We consider the coupled system for the state vector $\mathbf{v}(t) = [R(t), S_y(t), S_z(t)]^T$ with the coupling constant $J=1$. The equations of motion are:

\begin{align}
\frac{d R}{dt} &= \gamma S_z-\gamma(1-q)  R \\
\frac{d S_x}{dt} &= - \gamma S_x \\
\frac{d S_y}{dt} &= J S_z  - \gamma S_y \\
\frac{d S_z}{dt} &= -J S_y  - \gamma S_z + \gamma(1+q)R 
\end{align}

Given the initial conditions $S_y(0) = -1, S_z(0) = 0, R(0) = 1$, and $S_x(0) = 0$, it follows that $S_x(t) = 0$ for all $t \geq 0$.


Let $x = \lambda + \gamma$. The characteristic equation is $x^3 + ax^2 + bx + c = 0$ with:

\begin{align*}
a &= -\gamma q \\
b &= J^2 - \gamma^2(1+q) \\
c &= -\gamma q J^2
\end{align*}

Using Cardano's method, we define:

\begin{equation}
P = \frac{3b - a^2}{3}, \quad Q = \frac{2a^3 - 9ab + 27c}{27}
\end{equation}

The roots $x_j$ are derived from the intermediate terms $u$ and $v$:

\begin{equation}
u = \sqrt[3]{-\frac{Q}{2} + \sqrt{\frac{Q^2}{4} + \frac{P^3}{27}}}, \quad v = \sqrt[3]{-\frac{Q}{2} - \sqrt{\frac{Q^2}{4} + \frac{P^3}{27}}}
\end{equation}
The coefficients $c_j$ are calculated to satisfy the initial state $\mathbf{v}(0) = [1, -1, 0]^T$. The modified formula for $c_j$ (where the sign of the $S_y$ contribution is flipped) is:
\begin{equation}
c_j^{-} = \frac{x_k x_l - \gamma q(x_k + x_l) + \gamma^2 q - \frac{\gamma}{J}x_k x_l}{\gamma(x_j - x_k)(x_j - x_l)}
\end{equation}
where $(j, k, l)$ are cyclic permutations of $(1, 2, 3)$. 
The final exact solutions are:
\begin{align}
R^{-}(t) &= \gamma e^{-\gamma t} \sum_{j=1}^3 c_j^{-} x_j e^{x_j t} \\
S_y^{-}(t) &= J e^{-\gamma t} \sum_{j=1}^3 c_j^{-} (x_j - \gamma q) e^{x_j t} \\
S_z^{-}(t) &= e^{-\gamma t} \sum_{j=1}^3 c_j^{-} x_j (x_j - \gamma q) e^{x_j t}
\end{align}
\begin{equation} s_y^{-}(t) = \frac{J \sum_{j=1}^3 c_j^{-} (x_j - \gamma q) e^{x_j t}}{\gamma \sum_{j=1}^3 c_j^{-} x_j e^{x_j t}} \end{equation}
\begin{equation} 
s_z^{-}(t) = \frac{\sum_{j=1}^3 c_j^{-} x_j (x_j - \gamma q) e^{x_j t}}{\gamma \sum_{j=1}^3 c_j^{-} x_j e^{x_j t}} 
\end{equation}
Subsequently $K_3$ becomes,
\begin{equation}
K_3 = {s_y^{+}(t)} + \frac{1}{2}\Big({s_y^{+}(t)}- s_y^{-}(t)\Big) +\frac{1}{2}{s_y^{+}(t)}\Big({s_y^{+}(t)} + {s_y^{-}(t)}\Big) - {s_y^{+}(2t)}
\end{equation}
where, 
\begin{equation*} s_y^{+}(t) = \frac{J \sum_{j=1}^3 c_j^{+} (x_j - \gamma q) e^{x_j t}}{\gamma \sum_{j=1}^3 c_j^{+} x_j e^{x_j t}}, \end{equation*}
\begin{equation*}
s_y^{+}(2t) = \frac{J \sum_{j=1}^3 c_j^{+} (x_j - \gamma q) e^{2 x_j t}}{\gamma \sum_{j=1}^3 c_j^{+} x_j e^{2 x_j t}} 
\end{equation*}
\begin{equation*} s_y^{-}(t) = \frac{J \sum_{j=1}^3 c_j^{-} (x_j - \gamma q) e^{x_j t}}{\gamma \sum_{j=1}^3 c_j^{-} x_j e^{x_j t}} \end{equation*}
\section{Liouvillian Spectrum and Exceptional Points}
\label{app:liouvillian_spectrum}
\noindent This appendix summarizes the spectral structure of the hybrid Liouvillian and
the conditions under which exceptional points (EPs) arise.
For $\theta=\pi/2$, the hybrid Liouvillian master equation
\begin{equation}
\dot\rho=-i[H,\rho]
+2\gamma\Big(q\,L\rho L^\dagger-\tfrac12\{L^\dagger L,\rho\}\Big),
\end{equation}
with $H=-(J/2)\sigma_x$ and $L=\ket{\uparrow}\bra{\downarrow}$, can be written in
vectorized form as
\begin{equation}
\frac{d}{dt}|\rho\rangle\rangle=\mathcal{L}|\rho\rangle\rangle,
\end{equation}
using the ordering
$|\rho\rangle\rangle=(\rho_{00},\rho_{01},\rho_{10},\rho_{11})^{\mathrm T}$.
The corresponding Liouvillian superoperator reads
\begin{equation}
\mathcal{L}=
\begin{pmatrix}
0 & i\frac{J}{2} & -i\frac{J}{2} & 2q\gamma \\
i\frac{J}{2} & -\gamma & 0 & -i\frac{J}{2} \\
- i\frac{J}{2} & 0 & -\gamma & i\frac{J}{2} \\
0 & -i\frac{J}{2} & i\frac{J}{2} & -2\gamma
\end{pmatrix}.
\label{eq:L_matrix}
\end{equation}
The eigenvalues $\lambda$ are obtained from
$\det(\mathcal{L}-\lambda\mathbb I)=0$. Introducing the dimensionless variables
\begin{equation}
r\equiv\frac{\gamma}{J},
\qquad
x\equiv\frac{\lambda}{J},
\end{equation}
the characteristic polynomial factorizes as
\begin{equation}
\det(\mathcal{L}-\lambda\mathbb I)
=(\lambda+\gamma)\,P_3(\lambda),
\end{equation}
implying one exact eigenvalue
\begin{equation}
\lambda_1=-\gamma,
\end{equation}
while the remaining three eigenvalues are determined by the cubic equation
\begin{equation}
x^3+3r x^2+(2r^2+1)x+r(1-q)=0.
\label{eq:cubic}
\end{equation}
Exceptional points correspond to the coalescence of eigenvalues and eigenvectors
and occur when the cubic equation \eqref{eq:cubic} has multiple roots. This
condition is encoded in the vanishing of its discriminant,
\begin{equation}
\Delta
=4(r^2-1)^3-27q^2 r^2,
\end{equation}
which defines the EP locus in the parameter space as
\begin{equation}
\boxed{\,4(r^2-1)^3=27q^2 r^2\,}.
\label{eq:EP_condition}
\end{equation}
For generic $(r,q)$ satisfying Eq.~\eqref{eq:EP_condition}, the Liouvillian
exhibits a second-order EP characterized by a $2\times2$ Jordan block and
non-exponential (polynomially modified) temporal evolution. In the limit
$q=0$, the root $x=-r$ coincides with $\lambda_1=-\gamma$, and for $r^2=1$ this
leads to a triple degeneracy associated with a $3\times3$ Jordan block. The
condition \eqref{eq:EP_condition} therefore defines the boundary between
qualitatively distinct dynamical regimes of the hybrid evolution and underlies
the anomalous temporal behavior discussed in the main text.

\section{Compact Polynomial Representation of the Universal Fit Function}
\label{app:fit_parameter}
\noindent The maximal Leggett--Garg parameter is approximated using the high-accuracy
universal fit
\begin{equation}
    K_3^{\max}(\gamma,q)
    = A(\gamma)\,
      \tanh\!\big[B(\gamma)\,\log q + C(\gamma)\big]
      + D(\gamma),
\end{equation}
where each coefficient function is a 20th-order polynomial in $\gamma$.
For clarity, we write them in symbolic summation form:
\begin{align}
    A(\gamma) &= \sum_{n=0}^{20} a_n\,\gamma^{n}, 
    \quad \quad B(\gamma) = \sum_{n=0}^{20} b_n\,\gamma^{n}, \\
    C(\gamma) &= \sum_{n=0}^{20} c_n\,\gamma^{n}, 
    \quad \quad D(\gamma) = \sum_{n=0}^{20} d_n\,\gamma^{n}.
\end{align}

The numerical values of the coefficients 
$\{a_n\}$, $\{b_n\}$, $\{c_n\}$, and $\{d_n\}$  
are collected in Table~\ref{tab:coeffs}.  
These values originate from a global fit of order $20$ in~$\gamma$.

\begin{table*}[t]
    \centering
    \begin{tabular}{c|c|c|c|c}
        \hline\hline
        $n$ & $a_n$ & $b_n$ & $c_n$ & $d_n$ \\
        \hline
        20 & $+1.1125\times 10^{2}$ & $+2.1707\times 10^{2}$ & $+1.6410\times 10^{2}$ & $-7.3519\times 10^{1}$ \\
        19 & $-1.3767\times 10^{2}$ & $-1.8439\times 10^{2}$ & $+3.9330\times 10^{2}$ & $+1.1761\times 10^{2}$ \\
        18 & $-9.1780\times 10^{1}$ & $-1.5270\times 10^{2}$ & $-3.6510\times 10^{2}$ & $-1.7906\times 10^{1}$ \\
        17 & $+2.7423\times 10^{1}$ & $-8.2974\times 10^{1}$ & $-7.7988\times 10^{2}$ & $+4.1577\times 10^{1}$ \\
        16 & $+9.6074\times 10^{1}$ & $-3.1101\times 10^{1}$ & $-4.4073\times 10^{2}$ & $-4.8891\times 10^{1}$ \\
        15 & $+7.9796\times 10^{1}$ & $+2.6833\times 10^{2}$ & $+2.4773\times 10^{2}$ & $-5.7441\times 10^{1}$ \\
        14 & $+1.1570\times 10^{1}$ & $+1.8094\times 10^{2}$ & $+9.5807\times 10^{2}$ & $-2.0942\times 10^{1}$ \\
        13 & $-5.8861\times 10^{1}$ & $-6.9937\times 10^{1}$ & $+9.7489\times 10^{2}$ & $+2.0108$ \\
        12 & $-7.6905\times 10^{1}$ & $-9.0190\times 10^{1}$ & $-1.3577\times 10^{2}$ & $+4.8539\times 10^{1}$ \\
        11 & $-3.4865\times 10^{1}$ & $-6.9025\times 10^{1}$ & $-7.0012\times 10^{2}$ & $+7.2005\times 10^{1}$ \\
        10 & $+2.9117\times 10^{1}$ & $-5.0256\times 10^{1}$ & $-8.8019\times 10^{2}$ & $+1.6402\times 10^{1}$ \\
         9 & $+6.0866\times 10^{1}$ & $-4.2615\times 10^{1}$ & $-6.3958\times 10^{2}$ & $-6.9517\times 10^{1}$ \\
         8 & $+3.3930\times 10^{1}$ & $+5.8249\times 10^{1}$ & $+6.5620\times 10^{2}$ & $-9.2305\times 10^{1}$ \\
         7 & $-1.8958\times 10^{1}$ & $+1.5617\times 10^{2}$ & $+1.8993\times 10^{3}$ & $+2.6552\times 10^{1}$ \\
         6 & $-5.3469\times 10^{1}$ & $-8.1141\times 10^{1}$ & $-9.1085\times 10^{2}$ & $+1.2943\times 10^{2}$ \\
         5 & $-3.5672\times 10^{1}$ & $-6.9839\times 10^{1}$ & $-1.6438\times 10^{3}$ & $-4.6756\times 10^{1}$ \\
         4 & $+1.2011\times 10^{2}$ & $+5.0647\times 10^{1}$ & $+1.8487\times 10^{3}$ & $-7.9859\times 10^{1}$ \\
         3 & $-8.5770\times 10^{1}$ & $-5.1790$           & $-8.1818\times 10^{2}$ & $+7.6357\times 10^{1}$ \\
         2 & $+2.8757\times 10^{1}$ & $-2.1947$           & $+1.9281\times 10^{2}$ & $-2.7083\times 10^{1}$ \\
         1 & $-4.5638$              & $+1.9172\times 10^{-1}$ & $-2.9259\times 10^{1}$ & $+4.8657$ \\
         0 & $+6.2848\times 10^{-1}$ & $-4.9415\times 10^{-1}$ & $+1.7521$ & $+8.7271\times 10^{-1}$ \\
        \hline\hline
    \end{tabular}
    \caption{Polynomial coefficients for 
    $A(\gamma)$, $B(\gamma)$, $C(\gamma)$, and $D(\gamma)$ 
    appearing in the universal fit.  
    Each row lists the coefficient multiplying $\gamma^{n}$.}
    \label{tab:coeffs}
\end{table*}


\section{NSIT and AoT Conditions}
\label{app:NSIT_AoT}

This section derives the necessary tools for checking the status of violation of NSIT and AoT conditions. Utilizing  the hybrid dynamical map $\mathcal{E}_t$ generated by Eq.~(\ref{E01}), the general global three time joint probability is calculated as follows,
\begin{equation}
P(q_0,q_1,q_2) = \Tr(P_{q_2} \tilde{\rho}_{q_1}(t)) \Tr(P_{q_1} \tilde{\rho}_{q_0}(t)) \Tr(P_{q_0} {\rho}(0))
\end{equation}
while the general two time joint ($t_0 ,t_1)$ probability is:
\begin{equation}
P(q_0,q_1) = \Tr(P_{q_1} \tilde{\rho}_{q_0}(t)) \Tr(P_{q_0} {\rho}(0))
\end{equation}
and the general two time joint ($t_0, t_2)$ probability is:
\begin{equation}
P(q_0,q_2) = \Tr(P_{q_2} \tilde{\rho}_{q_0}(2t)) \Tr(P_{q_0} {\rho}(0))
\end{equation}
and general two time joint ($t_1, t_2)$ probability is:
\begin{equation}
P(q_1,q_2) = \Tr(P_{q_2} \tilde{\rho}_{q_1}(t)) \Tr(P_{q_1} \tilde{\rho}(t))
\end{equation}
The single time ($t_i$) probabilities are:
\begin{equation*}
P(q_0) = \Tr(P_{q_0} \tilde{\rho}(0))
\end{equation*}
\begin{equation*}
P(q_0) = \Tr(P_{q_1} \tilde{\rho}(t))
\end{equation*}
\begin{equation}
P(q_0) = \Tr(P_{q_2} \tilde{\rho}(2t))
\end{equation}
Here the $q_i \in \{+1,-1\}$ ($i\in \{0,1,2\}$) represent the outcome of the dichotomic observable $\sigma_y$ defined in main text and the $q_i$'s are measured at time $t_i$'s instances with $t_0 = 0, t_1 = t, t_2 = 2t$. Also we defined  $\tilde{\rho}(t) = \mathcal{E}_t(\rho(0) / \Tr[\mathcal{E}_{t}(\rho(0)]$, $\tilde{\rho}(2t) = \mathcal{E}_{2t}(\rho(0) / \Tr[\mathcal{E}_{2t}(\rho(0)]$ and $\tilde{\rho}_{q_i}(t) = {\mathcal{E}_t(P_{q_i})}/{\Tr[\mathcal{E}_t(P_{q_i})]}$ and $\tilde{\rho}_{q_i}(2t) = {\mathcal{E}_{2t}(P_{q_i})}/{\Tr[\mathcal{E}_{2t}(P_{q_i})]}$ (also provided in main text).

\begin{comment}

\documentclass[pra,aps,amsmath,reprint,10pt,twocolumn]{revtex4-2}

\usepackage{natbib}       
\usepackage{graphicx}     
\usepackage{xcolor}       
\usepackage{amsfonts}     
\usepackage{amssymb}      
\usepackage{amsthm}       
\usepackage{bbold}        
\usepackage{bbm}          
\usepackage{hyperref}     
\usepackage[normalem]{ulem}
\hypersetup{
    colorlinks=true,
    linkcolor=blue,
    citecolor=blue,
    filecolor=magenta,
    urlcolor=cyan,
}

\DeclareMathOperator{\Tr}{Tr}
\newcommand{\ket}[1]{\vert #1 \rangle}
\newcommand{\bra}[1]{\langle #1 \vert}
\newcommand{\braket}[2]{\langle #1 \vert #2 \rangle}

\begin{document}

\title{
\textcolor{black}{Leggett--Garg Violations in Hybrid Liouvillian Dynamics:\\ The Nonlinear Role of Detector Efficiency}}

\newcommand*{\affaddr}[1]{#1}
\newcommand*{\affmark}[1][*]{\textsuperscript{#1}}

\author{Sourav Paul\affmark[1]}
\email{sp20rs034@iiserkol.ac.in}
\author{Parveen Kumar \affmark[2]}
\email{parveenkumar622@gmail.com}
\author{Sourin Das \affmark[1]}
\email{sdas.du@gmail.com;sourin@iiserkol.ac.in}

\affiliation{
\affaddr{\affmark[1] Indian Institute of Science Education and Research Kolkata, Mohanpur, Nadia 741246, West Bengal, India}\\
\affaddr{\affmark[2] Department of Physics, Indian Institute of Technology Jammu, Jammu 181221, India}}
\begin{abstract}
We investigate violations of the Leggett--Garg inequality (LGI) in an open two-level system governed by a time-local hybrid Liouvillian. By introducing a continuous tuning parameter $q \in [0,1]$ that strictly quantifies detector efficiency---the probability of successfully recording and conditionally retaining environmental quantum jumps---we seamlessly interpolate between trace-preserving Lindblad dissipation ($q=1$) and non-Hermitian ``no-jump'' dynamics ($q=0$). We demonstrate that varying this detector efficiency induces highly nonlinear, logarithmic-scale amplifications in the LGI parameter $K_3$. While recent discrete map-based protocols rely on non-divisible superpositions of unitaries to transcend the temporal Tsirelson bound, our framework achieves extreme LGI violations approaching the algebraic maximum ($K_3 \to 3$) entirely within continuous, divisible quantum trajectory dynamics, revealing that macroscopic realism violations are acutely sensitive to the physical limits of measurement efficiency.
\end{abstract}
\maketitle

\section{Introduction}
\label{sec:I}
The boundary between classical and quantum behavior remains a central question in modern physics~\cite{Leggett2002,Schlosshauer2005,Zurek2003,Kofler2007}. While microscopic systems exhibit superposition and entanglement, macroscopic objects appear to possess definite properties independent of observation. Leggett and Garg formalized this tension through inequalities designed to test the compatibility of quantum mechanics with \textit{macrorealism} (MR) and \textit{non-invasive measurability} (NIM)~\cite{LG1985,Leggett1988}. Violations of the Leggett--Garg inequality (LGI) thus provide an operational witness of non-classical temporal correlations, analogous to Bell inequality violations for spatial correlations~\cite{Bell1964,Brunner2014,Emary2014}.

Experimentally, LGIs are probed by sequentially measuring a dichotomic observable $\hat{Q} = \pm 1$ at times $t_1, t_2, t_3$, from which the temporal correlators $C_{ij} = \langle \hat{Q}(t_i)\hat{Q}(t_j)\rangle$ are constructed. These define the LG parameter $K_3 = C_{12} + C_{23} - C_{13}$, which under macrorealism satisfies $-3 \le K_3 \le 1$~\cite{LG1985,Emary2014}. Violations of this bound have been observed in diverse platforms, including superconducting circuits~\cite{Palacios2010,Groblacher2015}, nuclear magnetic resonance systems~\cite{Katiyar2013,Mouse2017}, and nitrogen-vacancy centers in diamond~\cite{Waldherr2011}, confirming the robustness of LGI violation as a signature of temporal quantum coherence.

Within standard quantum mechanics, however, the magnitude of violation is constrained. For a two-level system undergoing unitary evolution with projective measurements, the LGI parameter is bounded by the temporal Tsirelson bound (TTB), also known as the Lüders bound, $(K_3)_{\rm QM} \le 1.5$~\cite{Fritz2010,Budiyono2013,Maroney2014}. This limit arises from the linearity of quantum evolution, state convexity, and the projective measurement update rule. Surpassing this bound requires modifying at least one of the assumptions of standard Hermitian, trace-preserving dynamics. Understanding how temporal quantum correlations behave in the presence of dissipation and measurement back-action is therefore an important problem in the theory of open quantum systems.

Several theoretical approaches have been proposed to exceed the Lüders bound, including non-Hermitian and $\mathcal{PT}$-symmetric Hamiltonians~\cite{Bender2007,Tang2018}, weak and generalized measurement schemes~\cite{Aharonov1988,Dressel2014,Knee2016}, and post-selected dynamics~\cite{Jordan2015}.

Despite these advances, a fundamental problem remains: the interpolation between the idealized non-Hermitian limit and the Lindblad limit of open quantum systems. While non-Hermitian dynamics are known to produce extreme LGI violations, it is not yet clear if these signatures survive when the dynamics are governed by a full master equation including dissipation and decoherence. Specifically, the question of whether strong LGI violations persist in the transition from pure-state quantum trajectories to ensemble-averaged mixed-state dynamics remains a critical gap in our understanding of temporal correlations.

In this work, we address the problem of interpolating between dissipative and non-Hermitian regimes by adopting a route based on open quantum system theory. Specifically, we utilize the ``Hybrid Liouvillian'' framework popularized in studies of quantum trajectories and continuous monitoring~\cite{Minganti2019,Kumar2020,Kumar2021}. Starting from the standard Gorini--Kossakowski--Sudarshan--Lindblad (GKSL) master equation~\cite{Gorini1976,Lindblad1976,Breuer2002,Plenio1998}, we introduce a post-selection parameter $q \in [0,1]$ representing the conditional retention of quantum jump records. This parameter naturally arises in continuous measurement and quantum trajectory descriptions as a measure of detector efficiency~\cite{Carmichael1993,Wiseman2009,Jacobs2014}.

Physically, the $q$-parameter serves as a direct bridge between two distinct physical limits. In a standard open system ($q=1$), the environment acts as an unobserved sink for information. Every ``quantum jump''---such as the emission of a photon into the environment---is effectively lost to the laboratory, resulting in trace-preserving Lindblad evolution. However, in modern experimental setups, the environment can be monitored. If we employ a detector to catch these jumps, $q$ becomes the probability that our detector successfully records and retains an event. 

By conditionally retaining only specific measurement outcomes (post-selection), we alter the system's effective evolution. When $q=0$, we consider the ``null-measurement'' or ``no-jump'' scenario, where we only keep experimental runs where no emission was detected. This filters the state, leading to effectively non-trace-preserving, non-Hermitian Hamiltonian dynamics. By varying $q$, we can smoothly tune the dynamics from the full noise of a dissipative environment to the high-information regime of post-selected trajectories. The resulting generator combines the structure of a Lindblad dissipator with the non-Hermitian effective Hamiltonian associated with quantum-jump conditioning~\cite{Kumar2020,Kumar2021}, yielding a hybrid master equation:
\begin{equation}
\dot{\rho} = -i[H, \rho] - \gamma \{L^\dagger L, \rho\} + 2\gamma q L \rho L^\dagger.
\label{eq:hybrid_liouvillian}
\end{equation}
Here, the parameter $q$ acts as a physical ``dial'' controlling the nonlinearity of the evolution and the degree of information recovered from the environment.

By solving the corresponding Bloch equations and computing two-time correlation functions, we show that the LGI parameter $K_3$ depends sensitively on $q$. We demonstrate that the post-selection-induced nonlinearity enables violations exceeding the Lüders bound and, in the extreme limit, approaching the algebraic maximum $K_3 = 3$. These results clarify the conditions under which strong LGI violations survive the interpolation into open system dynamics, providing realistic pathways for probing extreme macrorealism violations in dissipative environments.

The structure of this paper is as follows: In Sec.~\ref{sec:2}, we introduce our theoretical model, defining the system Hamiltonian and the hybrid Liouvillian master equation. We then detail the formalism for calculating the two-time correlators and the LGI parameter in this non-trace-preserving context. In Sec.~\ref{sec:3}, we present and discuss our numerical results, focusing on the rich behavior of $K_3$ as a function of the post-selection parameter $q$. Finally, in Sec.~\ref{sec:4}, we summarize our findings and discuss their broader implications for understanding quantum measurement, non-classicality, and potential experimental verifications.
\begin{figure*}[t!]
\centering
\includegraphics[width=0.4\textwidth]{figs/Fig_1a.png}
\includegraphics[width=0.52\textwidth]{figs/Fig_1b.png}
\caption{ \textbf{Maximal violation of the LG parameter.} (a) The maximal value $K_3^{\max}$ is shown as a function of the post-selection parameter $q$ (displayed on a logarithmic scale) and the dissipation ratio $\gamma$. The colormap encodes the magnitude of $K_3^{\max}$, highlighting the parameter regions where the violation exceeds the L"uders bound. The green curve marks the boundary at $K_3 = 1.5$: the region below this line corresponds to super-quantum violation, while the region above corresponds to the quantum-allowed domain. The white dashed curve denotes the exceptional line of the hybrid Liouvillian [Eq.~(\ref{eq:hybrid_liouvillian})], with further discussion provided in Appendix~\ref{app:liouvillian_spectrum}. (b) The corresponding gradient of $K_3^{\max}$ over the same parameter plane illustrates the descent structure of the landscape. In both panels, $J=1$, and $K_3$ is optimized over the measurement time interval $t$.}  
\label{fig1:k3_vs_q_gamma_wide}
\end{figure*}

\section{Hybrid Liouvillian Model}
\label{sec:2}

We consider a single qubit whose dynamics are governed by a hybrid Liouvillian master equation of the form
\begin{equation}
\frac{d \rho}{dt} = -i [H, \rho]
+ 2\gamma \Big( q\, L \rho L^{\dagger}
- \frac{1}{2} \{ L^{\dagger} L, \rho \} \Big),
\label{eq:hybrid_liouvillian}
\end{equation}
where $\rho$ is the system density matrix, $H$ is the Hamiltonian, $L$ is a quantum jump operator, and $\gamma$ denotes the dissipation rate. The real parameter $q \in [0,1]$ controls the effective nature of the evolution. For $q=1$, Eq.~(\ref{eq:hybrid_liouvillian}) reduces to the standard Gorini--Kossakowski--Sudarshan--Lindblad (GKSL) master equation, generating completely positive trace-preserving dynamics. In contrast, for $q=0$ the recycling term in the dissipator is absent, resulting in a non-trace-preserving evolution described by an effective non-Hermitian Hamiltonian, $H_{\mathrm{eff}} = H - i\gamma L^{\dagger}L$. This limit corresponds to post-selecting quantum trajectories for which no jump (e.g., photon emission) occurs during the evolution.

For the specific model studied here, the Hamiltonian and jump operator are chosen as
\begin{equation}
H = -\frac{J}{2} \left( \sin\theta\, \sigma_x + \cos\theta\, \sigma_z \right),
\qquad
L = \ket{\uparrow}\bra{\downarrow} = \sigma_+,
\end{equation}
where $J$ sets the energy scale of the coherent dynamics, $\theta$ parametrizes the Hamiltonian orientation on the Bloch sphere, and $\sigma_i$ are the Pauli operators. The jump operator $L=\sigma_+$ induces transitions from the ground state $\ket{\downarrow}$ to the excited state $\ket{\uparrow}$, representing the dissipative process. For the remainder of this work, we focus primarily on the case $\theta=\pi/2$, for which the Hamiltonian simplifies to $H=-(J/2)\sigma_x$. Rather than expanding the master equation into explicit equations of motion for each density-matrix element, we note that the components $\rho_{ij}$ are derived directly from this setup and solved numerically to obtain the time-evolved state $\rho(t)$.

To test the Leggett--Garg inequality, we consider the dichotomic observable $Q = \sigma_y$. The system is initialized in an eigenstate of $Q$, namely
\begin{equation}
\rho(0) = \ket{+_y}\bra{+_y},
\qquad
\ket{+_y} = \frac{1}{\sqrt{2}}(\ket{\uparrow} + i\ket{\downarrow}).
\end{equation}
Because the dynamics for $q \neq 1$ are not trace preserving, physical expectation values must be evaluated using the normalized density matrix
\begin{equation}
\tilde{\rho}(t) = \frac{\rho(t)}{\Tr[\rho(t)]}.
\end{equation}
The two-time correlation functions $C_{ij}$ entering the Leggett--Garg parameter are computed using the standard sequential-measurement protocol. A projective measurement of $Q$ at time $t_i$ collapses the state, which subsequently evolves under Eq.~(\ref{eq:hybrid_liouvillian}) until time $t_j$, where the expectation value of $Q$ is evaluated. Repeating this procedure for all relevant pairs of measurement times yields the Leggett--Garg parameter $K_3$, with further details provided in Sec.~\ref{sec:2A}.

Importantly, the master equation Eq.~(\ref{eq:hybrid_liouvillian}) is not merely phenomenological; it arises naturally from a microscopic system-detector interaction model. By treating the environment as a sequence of ancilla detectors, the parameter $q$ emerges from a Kraus map formalism representing the efficiency of post-selection. Specifically, the dynamics can be viewed as an interpolation between unconditioned evolution (where measurement outcomes are discarded, $q=1$) and perfect post-selection (where ``jump'' outcomes are rejected, $q=0$). We provide the rigorous derivation linking the microscopic system-detector Hamiltonian to this effective hybrid Liouvillian via the partial trace and continuum limit of the Kraus map in Appendix~\ref{app:derivation1}.






\subsection{Leggett--Garg Formalism}
\label{sec:2A}

The Leggett--Garg inequality provides a mathematical framework to test the principles of macrorealism, which assumes that a system exists in a definite state at all times and that measurements are non-invasive. In quantum mechanics, these assumptions are frequently violated, and the degree of this violation is quantified using two-time correlation functions. In our framework, we consider a dichotomic observable $Q = \sigma_y$. A projective measurement of $Q$ at any time $t_k$ collapses the system onto the eigenstates $\ket{\pm_y}$ with the corresponding projectors $P_{\pm} = \ket{\pm_y}\bra{\pm_y}$. The system is initially prepared in a state $\rho_0$ at $t_0=0$ and evolves under the hybrid dynamical map $\mathcal{E}_t$ generated by Eq.~(\ref{eq:hybrid_liouvillian}).

For the case where $q \neq 1$, the evolution is non-trace-preserving. Consequently, physical expectation values must be evaluated using the normalized state $\tilde{\rho}(t) = \mathcal{E}_t(\rho_0) / \Tr[\mathcal{E}_t(\rho_0)]$. To rigorously define the Leggett--Garg parameter $K_3(t)$, we evaluate three distinct two-time correlation functions using standard sequential-measurement protocols. The initial correlation $C_{01}(t)$ represents the correlation between the initial preparation and the first measurement at $t_1=t$. The sequential correlation $C_{12}(t)$ accounts for the intermediate measurement at $t_1$ and the subsequent conditional evolution to $t_2=2t$. Finally, the unmeasured correlation $C_{02}(t)$ measures the correlation between $t_0$ and $t_2=2t$ in the absence of any intermediate measurement.
\begin{figure*}[t!]
\centering
\includegraphics[width=0.48\textwidth]{figs/Fig_2a.png}
\includegraphics[width=0.46\textwidth]{figs/Fig_2b.png}
\caption{ \textbf{One-dimensional dependence of $K_3^{\text{max}}$ on $\gamma$ and $q$.} (a) The behavior of $K_3^{\text{max}}$ is shown as a function of the dissipation ratio $\gamma$ for fixed values of the post-selection parameter $q$. Each curve corresponds to a different choice of $q$, illustrating how post-selection strength modulates the $\gamma$-dependence of the maximal LGI violation. (b) The variation of $K_3^{\text{max}}$ with respect to $q$ (displayed on a logarithmic scale) is plotted for fixed values of $\gamma$. Distinct curves represent different $\gamma$ values, demonstrating the sensitivity of the LGI violation to changes in the post-selection parameter across several orders of magnitude.}
\label{fig2:k3_vs_q_gamma_1d_slice}
\end{figure*}
\subsubsection{Joint Probabilities and Correlation}
To rigorously define the Leggett--Garg parameter $K_3(t)$, we evaluate the two-time correlation functions by decomposing them into joint probabilities of obtaining specific outcomes at different time intervals. In our quantum mechanical treatment, the joint probability $P(q_i, q_j)$ of obtaining outcome $q_i$ at time $t_1$ and $q_j$ at time $t_2$ is defined as the product of the probability of the first outcome, $p_{q_i}(t) = \Tr[P_{q_i} \tilde{\rho}(t)]$, and the conditional expectation of the second. This requires evaluating the normalized post-measurement state, 
\begin{equation}
\tilde{\rho}_{q_i}(t) = \frac{\mathcal{E}_t(P_{q_i})}{\Tr[\mathcal{E}_t(P_{q_i})]},
\end{equation}
which represents the system after the first measurement, having evolved for an additional duration $t$.

Using this probabilistic framework, the three required correlation functions $C_{ij}$ emerge naturally as expectation values of the outcome products. The initial correlation between the preparation and the first measurement at $t_1=t$ evaluates directly to $C_{01}(t) = \Tr[\sigma_y \tilde{\rho}(t)]$. For the sequential correlation, which accounts for the intermediate measurement at $t_1$ and the subsequent evolution to $t_2=2t$, we take the conditional average over both possible intermediate outcomes, yielding
\begin{equation}
C_{12}(t) = \Tr[\sigma_y \tilde{\rho}_{+}(t)] p_{+}(t) - \Tr[\sigma_y \tilde{\rho}_{-}(t)] p_{-}(t).
\end{equation}
Finally, the unmeasured correlation between the initial state at $t_0$ and the final time $t_2=2t$, assuming no intermediate measurement occurred, simplifies to $C_{02}(t) = \Tr[\sigma_y \tilde{\rho}(2t)]$.

The Leggett--Garg parameter is then defined as the linear combination of these correlations,
\begin{equation}
K_3(t) = C_{01}(t) + C_{12}(t) - C_{02}(t).
\end{equation}
For any macrorealistic theory, the condition $K_3 \leq 1$ must be satisfied. Violations of this bound ($K_3 > 1$) indicate the presence of quantum coherence and non-invasive measurability issues, which are significantly influenced by the post-selection parameter $q$. Figure~\ref{fig1:k3_vs_q_gamma_wide} illustrates the numerical result of the optimal value of $K_3$ in a comprehensive parameter scan.

\subsection{Bloch-Vector Representation}
\label{sec:2B}

The hybrid Liouvillian dynamics generated by Eq.~(\ref{eq:hybrid_liouvillian}) admits a transparent representation in terms of the Bloch vector. Following the formalism for non-Hermitian PT-symmetric dynamics~\cite{Varma2021}, we decompose the density matrix as
\begin{equation}
\rho(t) = \frac{R(t)}{2}\,\mathbb{I} + \frac{1}{2}\,\vec{S}(t)\cdot\vec{\sigma},
\label{eq:bloch_decomp}
\end{equation}
where $R(t) = \mathrm{Tr}[\rho(t)]$ denotes the time-dependent trace and $\vec{S}(t) = \mathrm{Tr}[\rho(t)\vec{\sigma}]$ is the unnormalized Bloch vector. The physical, normalized state is described by $\vec{s}(t)=\vec{S}(t)/R(t)$ whenever $R(t) \neq 0$.

Substituting this decomposition into the master equation yields a set of coupled differential equations governing the evolution of the trace and the Bloch vector components. A key feature of these equations is that the trace is conserved ($\dot{R}=0$) only in the Lindbladian limit $q=1$. For any $q \neq 1$, the trace dynamics couples back into the spin evolution. Because the physical state depends on the ratio of $\vec{S}(t)$ to $R(t)$, this coupling renders the effective dynamics of the normalized vector $\vec{s}(t)$ highly nonlinear. We present the evolution of the normalized Bloch vector for the specific case of Hamiltonian orientation $\theta=\pi/2$ in Figure~\ref{fig4:bloch_vector_evolve_plane}, with further mathematical details provided in Appendix~\ref{app:bloch_special} which also includes analytic expressions (with some approximations) of these Bloch spin vector components

\section{Results}
\label{sec:3}
We present the numerical results for the Leggett--Garg parameter 
\begin{equation}
    K_3 = C_{01} + C_{12} - C_{02},
\end{equation}
computed using the three-time measurement protocol at $t_0=0$, $t_1=t$, and $t_2=2t$. The hybrid Liouvillian master equation~(\ref{eq:hybrid_liouvillian}) is solved numerically, and correlation functions are evaluated following the procedure outlined in Sec.~\ref{sec:2A}. For each simulation, $K_3$ is optimized over the measurement interval $t$, the dissipation ratio $\gamma$, and the post-selection parameter $q$.

\textit{\underline{Limiting Cases: $q=0$ and $q=1$}}: \ The dynamics are anchored by two distinct physical limits. Figure~\ref{fig1:k3_vs_q_gamma_wide} demonstrates that for \textbf{$q=1$}, the dynamics reduces to standard \textit{Lindblad evolution}. In this regime, increasing $\gamma$ suppresses quantum coherence and drives $K_3$ toward the classical limit ($K_3=1$), strictly obeying the L\"uders bound ($K_3 = 3/2$).

Conversely, in the strong post-selection limit (\textbf{$q \ll 1$}), the evolution is conditioned on quantum-jump-free trajectories, effectively stabilizing coherent dynamics. Upon optimizing over the full parameter space, we find that in this regime, $K_3$ can approach its algebraic maximum, $K_3 \to 3$. These extreme violations occur within narrow windows of $\gamma$, where the effective non-Hermitian evolution qualitatively modifies the Bloch-sphere dynamics, allowing for large angular separations between consecutive measurements that are unattainable within trace-preserving dynamics.

\textit{\underline{Interpolation Regime}}: \ The joint dependence of $K_3^{\max}$ on $q$ and $\gamma$ reveals a smooth transition between these two extremes. One-dimensional cuts, shown in Fig.~\ref{fig2:k3_vs_q_gamma_1d_slice}, indicate that decreasing $q$ at fixed $\gamma$ produces a sharp crossover in $K_3$. This behavior is well-described by a $\tanh$-like function of $\log q$, reflecting the sensitivity of temporal correlations to post-selection.

To obtain a compact and accurate description of the numerical behavior, the maximal Leggett--Garg parameter is modeled using the universal fit:
\begin{equation}
    K_3^{\max}(\gamma,q) = A(\gamma) \tanh\big[B(\gamma)\log q + C(\gamma)\big] + D(\gamma),
\end{equation}
where the coefficient functions $A(\gamma)$, $B(\gamma)$, $C(\gamma)$, and $D(\gamma)$ are each represented by a twentieth-order polynomial in $\gamma$. This polynomial--$\tanh$ structure (see Fig.~\ref{fig3:Fitted_K3_curve}) faithfully reproduces the full numerical data and correctly recovers the limiting behaviors at $q=0$ and $q=1$, ensuring the internal consistency of the model.

\textit{\underline{{Connection with the Exceptional Boundary}}}:\ Further insight into the origin of the enhanced LGI violation is obtained from the Bloch vector dynamics (Fig.~\ref{fig4:bloch_vector_evolve_plane}). As $q$ is reduced, the trajectories exhibit a clear qualitative transition: from damped motion at $q=1$, to oscillatory mixed-state dynamics at intermediate $q$, and finally to strongly accelerated evolution as the system approaches the \textit{exceptional boundary} at $q \ll 1$.

In this near-boundary regime, geometry-altering motion allows for constructive contributions from all correlation terms. These results establish that imperfect post-selection within hybrid Liouvillian dynamics enables a dramatic amplification of temporal quantum correlations, allowing the LGI parameter to approach its absolute upper bound. The combination of numerical optimization, analytic fitting, and geometric interpretation offers a predictive framework for understanding extreme LGI violations beyond standard open-system dynamics.

\begin{figure}[h!]
\centering
\includegraphics[width=0.35\textwidth]{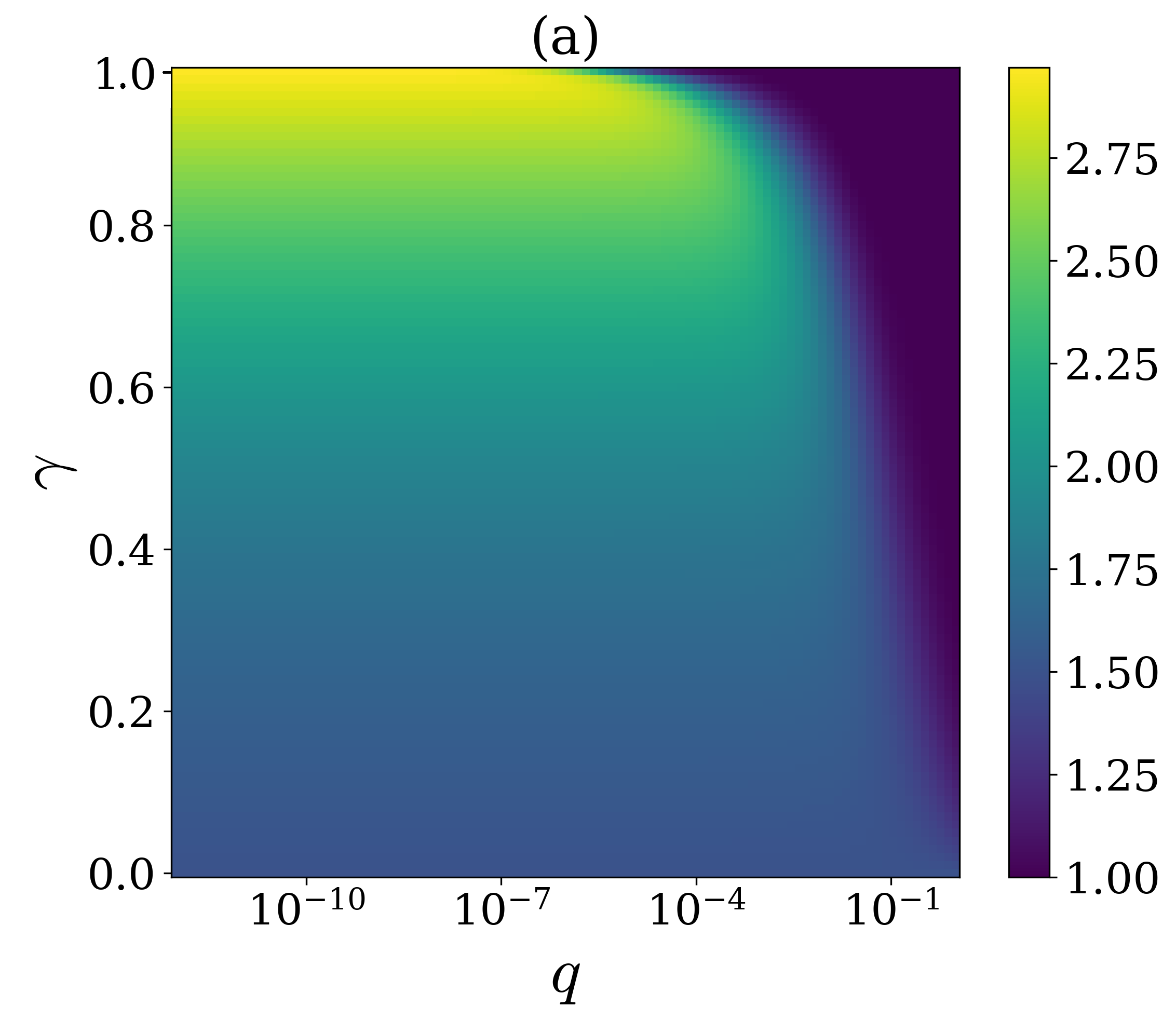}
\includegraphics[width=0.35\textwidth]{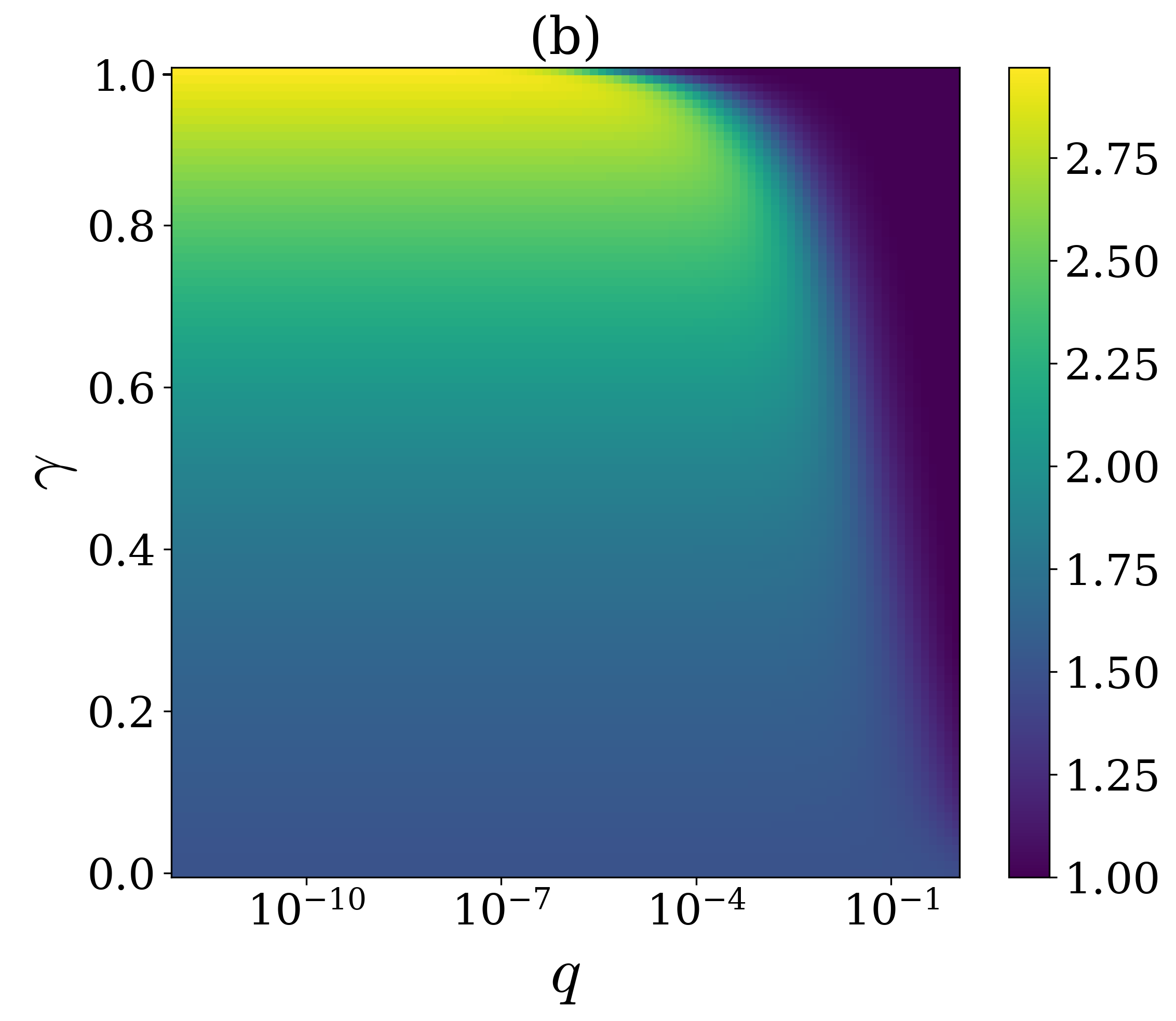}
\includegraphics[width=0.35\textwidth]{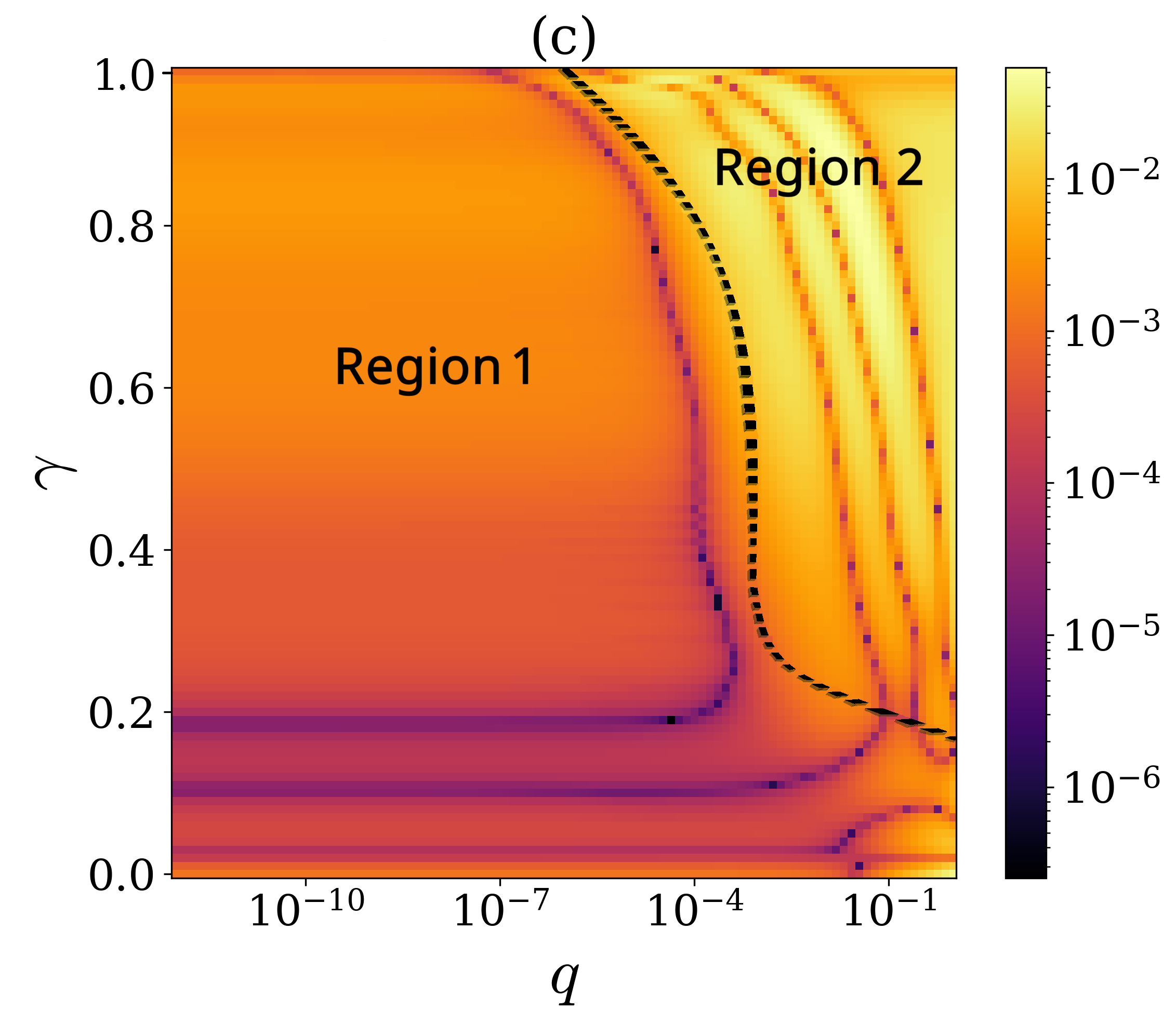}
\caption{ \textbf{Fitted curve for $K_3^{\text{max}}$.} (a) \ The top panel displays $K_3^{\text{max}}$ as a function of the parameters $\gamma$ and $q$ (with $q$ shown on a logarithmic scale). The interval $1 \leq \gamma \leq 2$ is not included, as this range exhibits no notable structure relevant for the fitting. (b) \ The lower panel shows the resulting fit, where $K_3^{\text{max}}(q,\gamma)$ is modeled as a polynomial function of $\gamma$ (up to twentieth order) combined with a $\tanh$-dependence on $\log q$ (see Appendix~\ref{app:fit_parameter} for more details). This composite form captures the rapid variation in $q$ and the smooth nonlinear dependence on $\gamma$.  (c) \ The accuracy of the fit is illustrated in the bottom residual plot. The explicit fitted expression is provided in the main text.}
\label{fig3:Fitted_K3_curve}
\end{figure}

\begin{figure*}[t!]
\includegraphics[width=0.32\textwidth]{figs/Polar_plot_q0.png}
\includegraphics[width=0.32\textwidth]{figs/Polar_plot_q105.png}
\includegraphics[width=0.32\textwidth]{figs/Polar_plot_q103.png}
\includegraphics[width=0.32\textwidth]{figs/Polar_plot_q102.png}
\includegraphics[width=0.32\textwidth]{figs/Polar_plot_q101.png}
\includegraphics[width=0.32\textwidth]{figs/Polar_plot_q1.png}
\caption{\textbf{Bloch vector trajectories under hybrid Liouvillian dynamics.} 
Evolution of the quantum state projected onto the $\hat{y}$-$\hat{z}$ plane, initialized in the $\vert +\rangle_y$ state. 
System parameters are fixed at $J=1$, $\gamma = 0.9905$, and $\theta = \pi/2$. 
Black and red markers indicate the initial ($t=0$) and asymptotic ($t \to \infty$) states, respectively. 
Blue markers trace the intermediate temporal evolution, while green arrows depict the instantaneous flow direction of the trajectory. 
Subplots correspond to varying parameter values arranged from top-left to bottom-right: $q \in \{0, 10^{-5}, 10^{-3}, 10^{-2}, 10^{-1}, 1\}$. 
A distinct qualitative transition in the dynamical topology is observable as $q$ increases.}
\label{fig4:bloch_vector_evolve_plane}
\end{figure*}

\section{Discussion and Conclusion}
\label{sec:4}

We have analyzed temporal quantum correlations in a driven two-level system governed by a hybrid Liouvillian master equation incorporating a tunable post-selection parameter \(q\). By optimizing the Leggett--Garg parameter~\cite{LG1985}
\[
K_3 = C_{01} + C_{12} - C_{02},
\]
over the measurement interval, dissipation strength \(\gamma\), and \(q\), we demonstrate that imperfect post-selection can strongly enhance violations of the Leggett--Garg inequality (LGI). In the trace-preserving limit \(q=1\), the dynamics reduces to standard Lindblad evolution~\cite{Lindblad1976,Gorini1976}, where dissipation suppresses coherence and drives \(K_3\) toward its classical bound. As \(q\) is reduced, the dynamics becomes conditioned on no-jump trajectories, effectively stabilizing coherence and enabling a substantial amplification of temporal quantum correlations.

A key result of this work is the pronounced sensitivity of \(K_3\) to the post-selection parameter \(q\). For fixed dissipation, \(K_3\) exhibits a sharp, approximately logarithmic dependence on \(q\), implying that even modest deviations from perfect post-selection can significantly alter the magnitude of LGI violations. While extreme violations require strong trajectory conditioning, substantial violations persist over a broad range of \(q \neq 1\), underscoring the experimental relevance of the hybrid Liouvillian framework.

Full optimization reveals that near-algebraic values of \(K_3\) arise only within narrow regions of intermediate dissipation and strong post-selection (\(q \ll 1\)). In this regime, the conditioned dynamics is effectively non-Hermitian~\cite{Rotter2009,Daley2014}, leading to accelerated evolution on the Bloch sphere. The resulting large angular separations between successive measurement times allow constructive interference between correlation functions, driving \(K_3\) far beyond the Lüders bound of \(1.5\) for unitary qubit dynamics~\cite{Luders1951,Budroni2014}---a regime inaccessible within conventional trace-preserving Lindblad evolution.

To contextualize our framework, it is instructive to contrast our approach with preceding proposals centered on the superposition of unitary maps—most notably the work of Mahesh and Usha Devi~\cite{MaheshUshaDevi2024}. This research direction explored extending the principle of superposition beyond individual quantum states to the broader domain of quantum dynamical maps.

Central to this architecture is the construction of a coherent linear combination of unitary propagators, formulated as:
\begin{equation}
    A(t) = c_1 U_1(t) + c_2 U_2(t).
\end{equation}
Such a construction defines a map that is, in general, both non-unitary and non-divisible. Within this framework, violations transcending the temporal Tsirelson bound emerge directly from the coherent superposition of distinct unitary evolutions. This synthesis results in intrinsically non-divisible dynamics, fundamentally defined and manifested at the level of quantum maps. By contrast, the hybrid Liouvillian formalism retains a continuous-time, time-local master equation, with the enhancement of temporal correlations emerging from controlled post-selection rather than from map-level superposition. The existence of an optimal post-selection parameter maximizing \(K_3\) is a distinctive feature of our approach and has no direct analogue in superposed-unitary schemes.

The dependence of the maximal violation \(K_3^{\max}(q,\gamma)\) can be accurately captured by a compact analytic form combining a smooth polynomial function of \(\gamma\) with a \(\tanh\)-like dependence on \(\log q\). This parametrization faithfully reproduces the numerical results across the relevant parameter space and recovers the expected limits corresponding to purely trace-preserving and strongly post-selected dynamics.

In summary, we have shown that hybrid Liouvillian dynamics provides a simple yet powerful route to amplifying temporal quantum correlations in a two-level system. By tuning dissipation and imperfect post-selection, LGI violations approaching their algebraic maximum can be achieved within a time-local master-equation framework. Our results complement and extend earlier superposed-unitary approaches~\cite{MaheshUshaDevi2024} and establish post-selection as an effective and experimentally meaningful control parameter for exploring extreme temporal nonclassicality.

\section{Acknowledgement} 
S.P. offers his gratitude to the Council of Scientific and Industrial Research (CSIR), Govt. of India for financial support. PK acknowledges support from the IIT Jammu Initiation Grant No. SGT-100106.

\appendix

\section{Microscopic Derivation of the Hybrid Dynamics}
\label{app:derivation1}

In this appendix, we provide the microscopic foundation for the hybrid Liouvillian model utilized in the main text. We proceed by defining the system-detector interaction and deriving the continuous-time master equation by explicitly tracing out the detector degrees of freedom and incorporating imperfect post-selection.

\subsection{System-Detector Hamiltonian}

We consider a system coupled to a detector, modeled as a two-level quantum object. The system Hamiltonian is given by $H_s = \omega \hat{n} \cdot \boldsymbol{\sigma}$, where $\omega$ is the Zeeman energy, and $\hat{n}$ is a unit vector parametrized by spherical coordinates $(\theta, \phi)$. The detector is prepared in the initial state $\rho_d^0 = \frac{1}{2}(\mathbb{I} + \hat{m} \cdot \boldsymbol{\sigma})$, where $\hat{m}$ defines the initialization direction.

The joint system-detector state, initially $\rho(t) = \rho_s(t) \otimes \rho_d^0$, evolves under the interaction Hamiltonian
\begin{equation}
H_{s-d} = \sqrt{2}J [\sigma_x^s \otimes \sigma_x^d + \sigma_y^s \otimes \sigma_y^d ],
\end{equation}
where $J$ is the coupling strength. This anisotropic interaction couples the system's spin direction orthogonal to the detector's initialization. Following the interaction, the detector is measured projectively and reset.

\subsection{Kraus Representation and Imperfect Post-Selection}

The reduced dynamics of the system over an infinitesimal time step $\delta t$ is obtained by tracing out the detector degrees of freedom. Let $U(\delta t) = \exp[-i(H_s \otimes \mathbb{I} + H_{s-d})\delta t]$ be the joint unitary evolution operator. The updated system state is
\begin{equation}
\tilde{\rho}_s(t+\delta t) = \Tr_d \left[ U(\delta t) (\rho_s(t) \otimes \rho_d^0) U^\dagger(\delta t) \right].
\end{equation}
Evaluating the partial trace in the eigenbasis $\{|k\rangle_d\}$ of the detector allows us to express the evolution in terms of Kraus operators $M_k = {}_d\langle k | U(\delta t) | \psi_{init} \rangle_d$, such that $\tilde{\rho}_s(t+\delta t) = \sum_k M_k \rho_s(t) M_k^\dagger$.
For the specific interaction chosen, this procedure yields two dominant operators representing the "no-jump" ($M_0$) and "jump" ($M_1$) scenarios:
\begin{align}
M_0 &= \mathbb{I} - i H_{\text{eff}} \delta t, \\
M_1 &= \sqrt{2\gamma \delta t} \, L,
\end{align}
where $H_{\text{eff}} = H - i\gamma L^\dagger L$ and $L = \sigma_+$, with the effective rate $\gamma$ defined by the continuum scaling of the coupling $J$.

In a standard trace-preserving map ($q=1$), both outcomes are kept. To model imperfect post-selection, we introduce the efficiency parameter $q \in [0,1]$. This parameter scales the contribution of the "jump" trajectory, reflecting a scenario where a portion of the jump outcomes are discarded (post-selected out). The modified Kraus map for the unnormalized density matrix becomes:
\begin{equation}
\tilde{\rho}_{\text{S}}(t+\delta t) = M_0 \tilde{\rho}_{\text{S}}(t) M_0^{\dagger} + q M_1 \tilde{\rho}_{\text{S}}(t) M_1^{\dagger}.
\end{equation}
Substituting the explicit forms of $M_0$ and $M_1$:
\begin{align}
\tilde{\rho}_{\text{S}}(t+\delta t) &\approx (\mathbb{I} - i H_{\text{eff}} \delta t) \rho(t) (\mathbb{I} + i H_{\text{eff}}^\dagger \delta t) \nonumber \\
&\quad + 2\gamma q \delta t L \rho(t) L^\dagger.
\end{align}
Expanding to first order in $\delta t$, we obtain:
\begin{align}
\frac{\tilde{\rho}_{\text{S}}(t+\delta t) - \rho(t)}{\delta t} &\approx -i[H, \rho(t)] - \gamma \{L^\dagger L, \rho(t)\} \nonumber \\
&\quad + 2\gamma q L \rho(t) L^\dagger.
\end{align}
Taking the limit $\delta t \to 0$, we arrive at the hybrid Liouvillian master equation presented in Eq.~(\ref{eq:hybrid_liouvillian}). This derivation explicitly demonstrates that the term $2\gamma q L \rho L^\dagger$ arises from the partial retention of the jump trajectories, while the anti-commutator term remains unaffected, preserving the coherent non-Hermitian decay characteristic of the no-jump evolution.

\section{Bloch Dynamics at $\theta = \pi/2$}
\label{app:bloch_special}

In this appendix, we provide the explicit equations of motion for the specific parameter regime $\theta = \pi/2$ used in the main text. In this limit, the Hamiltonian reduces to $H = -(J/2)\sigma_x$, and the general equations derived in Sec.~\ref{sec:2B} simplify significantly. The evolution of the $S_x$ component decouples from the system:
\begin{equation}
\frac{d S_x}{dt} = -\gamma S_x.
\end{equation}
Given the initial state $\rho(0) = \ket{+_y}\bra{+_y}$, we have $S_x(0)=0$, which implies $S_x(t)=0$ for all $t$. The dynamics are therefore strictly confined to the $y$-$z$ plane.

The remaining coupled dynamics for the trace $R$ and the components $S_y, S_z$ can be written in compact matrix form as:
\begin{equation}
\frac{d}{dt}
\begin{pmatrix} R \\ S_y \\ S_z \end{pmatrix}
=
\begin{pmatrix}
-\gamma(1-q) & 0 & \gamma(1-q) \\
0 & -\gamma & J \\
\gamma(1+q) & -J & -\gamma(1+q)
\end{pmatrix}
\begin{pmatrix} R \\ S_y \\ S_z \end{pmatrix}
\label{EB2}
\end{equation}
This linear system describes the interplay between the coherent driving $J$ and the measurement-induced decay and recycling rates parameterized by $\gamma$ and $q$.
\subsection{Approx. Analytical Solution for the Coupled Spin System}
We rewrite the eq.(\ref{EB2}) with an approximation that we assume $ \gamma q S_z\simeq 0$ for the all equations and get the following form. {\it These following equations correctly represent the original one  numerically which we checked.}
\begin{align}
\frac{d R}{dt} &= \gamma S_z-\gamma(1-q)  R, \label{eq:bloch_R} \\
\frac{d S_y}{dt} &= J S_z  - \gamma S_y, \\
\frac{d S_z}{dt} &= -J S_y  - \gamma S_z + \gamma(1+q)R. \label{eq:bloch_Sz}
\end{align}
Below we present the analytical solution for the state vector $\mathbf{v}(t) = [R(t), S_y(t), S_z(t)]^T$ based on the coupled differential equations with $J=1$ and initial conditions:

\begin{equation}
R(0) = 1, \quad S_y(0) = 1, \quad  S_z(0) = 0
\end{equation}

The system of equations is solved using the matrix method for the state vector $\mathbf{v}(t) = [R(t), S_y(t), S_z(t)]^T$. Given the initial conditions $\mathbf{v}(0) = [1, 1, 0]^T$.

\subsection*{1.1 The Characteristic Equation}

The eigenvalues $\lambda$ are found by solving the cubic equation derived from $\det(A - \lambda I) = 0$. Let $x = \lambda + \gamma$. The values of $x$ are the roots of:

\begin{equation}
x^3 + ax^2 + bx + c = 0
\end{equation}

where the coefficients are defined by the physical parameters:

\begin{align*}
a &= -\gamma q \\
b &= J^2 - \gamma^2(1+q) \\
c &= -\gamma q J^2
\end{align*}

\subsection*{1.2 Eigenvalues via Cardano's Method}

Defining the depressed cubic parameters:

\begin{equation}
P = \frac{3b - a^2}{3}, \quad Q = \frac{2a^3 - 9ab + 27c}{27}
\end{equation}

The intermediate terms for the exact roots are:

\begin{equation}
u = \sqrt[3]{-\frac{Q}{2} + \sqrt{\frac{Q^2}{4} + \frac{P^3}{27}}}, \quad v = \sqrt[3]{-\frac{Q}{2} - \sqrt{\frac{Q^2}{4} + \frac{P^3}{27}}}
\end{equation}

The three roots $x_j$ (and thus $\lambda_j = x_j - \gamma$) are:

\begin{align*}
x_1 &= u + v - \frac{a}{3} \\
x_2 &= \omega u + \omega^2 v - \frac{a}{3} \\
x_3 &= \omega^2 u + \omega v - \frac{a}{3}
\end{align*}

where $\omega = e^{i 2\pi / 3} = -\frac{1}{2} + i\frac{\sqrt{3}}{2}$.

\subsection*{1.3 Time-Domain Solutions}

Using the eigenvectors $\mathbf{u}_j = [\gamma x_j, J(x_j - \gamma q), x_j(x_j - \gamma q)]^T$, the exact solutions are:

\begin{align}
R^{+}(t) &= \gamma e^{ - \gamma t}\sum_{j=1}^3 c_j^{+} x_j e^{x_j t} \\
S_y^{+}(t) &= J e^{ - \gamma t}\sum_{j=1}^3 c_j^{+} (x_j - \gamma q) e^{x_j t} \\
S_z^{+}(t) &= e^{ - \gamma t}\sum_{j=1}^3 c_j^{+} x_j (x_j - \gamma q) e^{qx_j t}
\end{align}

The coefficients $c_j$ are determined by the initial conditions:

\begin{equation}
c_j^{+} = \frac{x_k x_l - \gamma q(x_k + x_l) + \gamma^2 q + \frac{\gamma}{J}x_k x_l}{\gamma(x_j - x_k)(x_j - x_l)}
\end{equation}

for $(j, k, l)$ as cyclic permutations of $(1, 2, 3)$.

\begin{equation} s_y^{+}(t) = \frac{J \sum_{j=1}^3 c_j^{+} (x_j - \gamma q) e^{x_j t}}{\gamma \sum_{j=1}^3 c_j^{+} x_j e^{x_j t}} \end{equation}

\begin{equation} 
s_z^{+}(t) = \frac{\sum_{j=1}^3 c_j^{+} x_j (x_j - \gamma q) e^{x_j t}}{\gamma \sum_{j=1}^3 c_j^{+} x_j e^{x_j t}} 
\end{equation}


We consider the coupled system for the state vector $\mathbf{v}(t) = [R(t), S_y(t), S_z(t)]^T$ with the coupling constant $J=1$. The equations of motion are:

\begin{align}
\frac{d R}{dt} &= \gamma S_z-\gamma(1-q)  R \\
\frac{d S_x}{dt} &= - \gamma S_x \\
\frac{d S_y}{dt} &= J S_z  - \gamma S_y \\
\frac{d S_z}{dt} &= -J S_y  - \gamma S_z + \gamma(1+q)R 
\end{align}

Given the initial conditions $S_y(0) = -1, S_z(0) = 0, R(0) = 1$, and $S_x(0) = 0$, it follows that $S_x(t) = 0$ for all $t \geq 0$.


Let $x = \lambda + \gamma$. The characteristic equation is $x^3 + ax^2 + bx + c = 0$ with:

\begin{align*}
a &= -\gamma q \\
b &= J^2 - \gamma^2(1+q) \\
c &= -\gamma q J^2
\end{align*}

Using Cardano's method, we define:

\begin{equation}
P = \frac{3b - a^2}{3}, \quad Q = \frac{2a^3 - 9ab + 27c}{27}
\end{equation}

The roots $x_j$ are derived from the intermediate terms $u$ and $v$:

\begin{equation}
u = \sqrt[3]{-\frac{Q}{2} + \sqrt{\frac{Q^2}{4} + \frac{P^3}{27}}}, \quad v = \sqrt[3]{-\frac{Q}{2} - \sqrt{\frac{Q^2}{4} + \frac{P^3}{27}}}
\end{equation}


The coefficients $c_j$ are calculated to satisfy the initial state $\mathbf{v}(0) = [1, -1, 0]^T$. The modified formula for $c_j$ (where the sign of the $S_y$ contribution is flipped) is:

\begin{equation}
c_j^{-} = \frac{x_k x_l - \gamma q(x_k + x_l) + \gamma^2 q - \frac{\gamma}{J}x_k x_l}{\gamma(x_j - x_k)(x_j - x_l)}
\end{equation}

where $(j, k, l)$ are cyclic permutations of $(1, 2, 3)$.

The final exact solutions are:

\begin{align}
R^{-}(t) &= \gamma e^{-\gamma t} \sum_{j=1}^3 c_j^{-} x_j e^{x_j t} \\
S_y^{-}(t) &= J e^{-\gamma t} \sum_{j=1}^3 c_j^{-} (x_j - \gamma q) e^{x_j t} \\
S_z^{-}(t) &= e^{-\gamma t} \sum_{j=1}^3 c_j^{-} x_j (x_j - \gamma q) e^{x_j t}
\end{align}

\begin{equation} s_y^{-}(t) = \frac{J \sum_{j=1}^3 c_j^{-} (x_j - \gamma q) e^{x_j t}}{\gamma \sum_{j=1}^3 c_j^{-} x_j e^{x_j t}} \end{equation}

\begin{equation} 
s_z^{-}(t) = \frac{\sum_{j=1}^3 c_j^{-} x_j (x_j - \gamma q) e^{x_j t}}{\gamma \sum_{j=1}^3 c_j^{-} x_j e^{x_j t}} 
\end{equation}

Subsequently $K_3$ becomes,

\begin{equation}
K_3 = {s_y^{+}(t)} + \frac{1}{2}\Big({s_y^{+}(t)}- s_y^{-}(t)\Big) +\frac{1}{2}{s_y^{+}(t)}\Big({s_y^{+}(t)} + {s_y^{-}(t)}\Big) - {s_y^{+}(2t)}
\end{equation}

where, 

\begin{equation*} s_y^{+}(t) = \frac{J \sum_{j=1}^3 c_j^{+} (x_j - \gamma q) e^{x_j t}}{\gamma \sum_{j=1}^3 c_j^{+} x_j e^{x_j t}}, \end{equation*}

\begin{equation*}
s_y^{+}(2t) = \frac{J \sum_{j=1}^3 c_j^{+} (x_j - \gamma q) e^{2 x_j t}}{\gamma \sum_{j=1}^3 c_j^{+} x_j e^{2 x_j t}} 
\end{equation*}

\begin{equation*} s_y^{-}(t) = \frac{J \sum_{j=1}^3 c_j^{-} (x_j - \gamma q) e^{x_j t}}{\gamma \sum_{j=1}^3 c_j^{-} x_j e^{x_j t}} \end{equation*}
\section{Liouvillian Spectrum and Exceptional Points}
\label{app:liouvillian_spectrum}

This appendix summarizes the spectral structure of the hybrid Liouvillian and
the conditions under which exceptional points (EPs) arise.

For $\theta=\pi/2$, the hybrid Liouvillian master equation
\begin{equation}
\dot\rho=-i[H,\rho]
+2\gamma\Big(q\,L\rho L^\dagger-\tfrac12\{L^\dagger L,\rho\}\Big),
\end{equation}
with $H=-(J/2)\sigma_x$ and $L=\ket{\uparrow}\bra{\downarrow}$, can be written in
vectorized form as
\begin{equation}
\frac{d}{dt}|\rho\rangle\rangle=\mathcal{L}|\rho\rangle\rangle,
\end{equation}
using the ordering
$|\rho\rangle\rangle=(\rho_{00},\rho_{01},\rho_{10},\rho_{11})^{\mathrm T}$.
The corresponding Liouvillian superoperator reads
\begin{equation}
\mathcal{L}=
\begin{pmatrix}
0 & i\frac{J}{2} & -i\frac{J}{2} & 2q\gamma \\
i\frac{J}{2} & -\gamma & 0 & -i\frac{J}{2} \\
- i\frac{J}{2} & 0 & -\gamma & i\frac{J}{2} \\
0 & -i\frac{J}{2} & i\frac{J}{2} & -2\gamma
\end{pmatrix}.
\label{eq:L_matrix}
\end{equation}

The eigenvalues $\lambda$ are obtained from
$\det(\mathcal{L}-\lambda\mathbb I)=0$. Introducing the dimensionless variables
\begin{equation}
r\equiv\frac{\gamma}{J},
\qquad
x\equiv\frac{\lambda}{J},
\end{equation}
the characteristic polynomial factorizes as
\begin{equation}
\det(\mathcal{L}-\lambda\mathbb I)
=(\lambda+\gamma)\,P_3(\lambda),
\end{equation}
implying one exact eigenvalue
\begin{equation}
\lambda_1=-\gamma,
\end{equation}
while the remaining three eigenvalues are determined by the cubic equation
\begin{equation}
x^3+3r x^2+(2r^2+1)x+r(1-q)=0.
\label{eq:cubic}
\end{equation}

Exceptional points correspond to the coalescence of eigenvalues and eigenvectors
and occur when the cubic equation \eqref{eq:cubic} has multiple roots. This
condition is encoded in the vanishing of its discriminant,
\begin{equation}
\Delta
=4(r^2-1)^3-27q^2 r^2,
\end{equation}
which defines the EP locus in the parameter space as
\begin{equation}
\boxed{\,4(r^2-1)^3=27q^2 r^2\,}.
\label{eq:EP_condition}
\end{equation}

For generic $(r,q)$ satisfying Eq.~\eqref{eq:EP_condition}, the Liouvillian
exhibits a second-order EP characterized by a $2\times2$ Jordan block and
non-exponential (polynomially modified) temporal evolution. In the limit
$q=0$, the root $x=-r$ coincides with $\lambda_1=-\gamma$, and for $r^2=1$ this
leads to a triple degeneracy associated with a $3\times3$ Jordan block. The
condition \eqref{eq:EP_condition} therefore defines the boundary between
qualitatively distinct dynamical regimes of the hybrid evolution and underlies
the anomalous temporal behavior discussed in the main text.

\section{Compact Polynomial Representation of the Universal Fit Function}
\label{app:fit_parameter}

The maximal Leggett--Garg parameter is approximated using the high-accuracy
universal fit
\begin{equation}
    K_3^{\max}(\gamma,q)
    = A(\gamma)\,
      \tanh\!\big[B(\gamma)\,\log q + C(\gamma)\big]
      + D(\gamma),
\end{equation}
where each coefficient function is a 20th-order polynomial in $\gamma$.
For clarity, we write them in symbolic summation form:
\begin{align}
    A(\gamma) &= \sum_{n=0}^{20} a_n\,\gamma^{n}, \\
    B(\gamma) &= \sum_{n=0}^{20} b_n\,\gamma^{n}, \\
    C(\gamma) &= \sum_{n=0}^{20} c_n\,\gamma^{n}, \\
    D(\gamma) &= \sum_{n=0}^{20} d_n\,\gamma^{n}.
\end{align}

The numerical values of the coefficients 
$\{a_n\}$, $\{b_n\}$, $\{c_n\}$, and $\{d_n\}$  
are collected in Table~\ref{tab:coeffs}.  
These values originate from a global fit of order $20$ in~$\gamma$.

\begin{table*}[t]
    \centering
    \caption{Polynomial coefficients for 
    $A(\gamma)$, $B(\gamma)$, $C(\gamma)$, and $D(\gamma)$ 
    appearing in the universal fit.  
    Each row lists the coefficient multiplying $\gamma^{n}$.}
    \label{tab:coeffs}
    \begin{tabular}{c|c|c|c|c}
        \hline\hline
        $n$ & $a_n$ & $b_n$ & $c_n$ & $d_n$ \\
        \hline
        20 & $+1.1125\times 10^{2}$ & $+2.1707\times 10^{2}$ & $+1.6410\times 10^{2}$ & $-7.3519\times 10^{1}$ \\
        19 & $-1.3767\times 10^{2}$ & $-1.8439\times 10^{2}$ & $+3.9330\times 10^{2}$ & $+1.1761\times 10^{2}$ \\
        18 & $-9.1780\times 10^{1}$ & $-1.5270\times 10^{2}$ & $-3.6510\times 10^{2}$ & $-1.7906\times 10^{1}$ \\
        17 & $+2.7423\times 10^{1}$ & $-8.2974\times 10^{1}$ & $-7.7988\times 10^{2}$ & $+4.1577\times 10^{1}$ \\
        16 & $+9.6074\times 10^{1}$ & $-3.1101\times 10^{1}$ & $-4.4073\times 10^{2}$ & $-4.8891\times 10^{1}$ \\
        15 & $+7.9796\times 10^{1}$ & $+2.6833\times 10^{2}$ & $+2.4773\times 10^{2}$ & $-5.7441\times 10^{1}$ \\
        14 & $+1.1570\times 10^{1}$ & $+1.8094\times 10^{2}$ & $+9.5807\times 10^{2}$ & $-2.0942\times 10^{1}$ \\
        13 & $-5.8861\times 10^{1}$ & $-6.9937\times 10^{1}$ & $+9.7489\times 10^{2}$ & $+2.0108$ \\
        12 & $-7.6905\times 10^{1}$ & $-9.0190\times 10^{1}$ & $-1.3577\times 10^{2}$ & $+4.8539\times 10^{1}$ \\
        11 & $-3.4865\times 10^{1}$ & $-6.9025\times 10^{1}$ & $-7.0012\times 10^{2}$ & $+7.2005\times 10^{1}$ \\
        10 & $+2.9117\times 10^{1}$ & $-5.0256\times 10^{1}$ & $-8.8019\times 10^{2}$ & $+1.6402\times 10^{1}$ \\
         9 & $+6.0866\times 10^{1}$ & $-4.2615\times 10^{1}$ & $-6.3958\times 10^{2}$ & $-6.9517\times 10^{1}$ \\
         8 & $+3.3930\times 10^{1}$ & $+5.8249\times 10^{1}$ & $+6.5620\times 10^{2}$ & $-9.2305\times 10^{1}$ \\
         7 & $-1.8958\times 10^{1}$ & $+1.5617\times 10^{2}$ & $+1.8993\times 10^{3}$ & $+2.6552\times 10^{1}$ \\
         6 & $-5.3469\times 10^{1}$ & $-8.1141\times 10^{1}$ & $-9.1085\times 10^{2}$ & $+1.2943\times 10^{2}$ \\
         5 & $-3.5672\times 10^{1}$ & $-6.9839\times 10^{1}$ & $-1.6438\times 10^{3}$ & $-4.6756\times 10^{1}$ \\
         4 & $+1.2011\times 10^{2}$ & $+5.0647\times 10^{1}$ & $+1.8487\times 10^{3}$ & $-7.9859\times 10^{1}$ \\
         3 & $-8.5770\times 10^{1}$ & $-5.1790$           & $-8.1818\times 10^{2}$ & $+7.6357\times 10^{1}$ \\
         2 & $+2.8757\times 10^{1}$ & $-2.1947$           & $+1.9281\times 10^{2}$ & $-2.7083\times 10^{1}$ \\
         1 & $-4.5638$              & $+1.9172\times 10^{-1}$ & $-2.9259\times 10^{1}$ & $+4.8657$ \\
         0 & $+6.2848\times 10^{-1}$ & $-4.9415\times 10^{-1}$ & $+1.7521$ & $+8.7271\times 10^{-1}$ \\
        \hline\hline
    \end{tabular}
\end{table*}

\section{NSIT and AoT Conditions}
\label{app:NSIT_AoT}

This appendix summarizes the status of noninvasive measurability (NSIT) and
arrow-of-time (AoT) conditions for the hybrid Liouvillian dynamics considered
in the main text, and clarifies their implications for macroscopic realism.

For dichotomic outcomes $m_i=\pm1$ measured at times $t_1<t_2<t_3$, macroscopic
realism is equivalent to the existence of a global joint probability
distribution $P(m_1,m_2,m_3)$. This, in turn, requires the simultaneous validity
of all NSIT and AoT constraints.

Explicit evaluation of multitime probabilities generated by the hybrid
Liouvillian shows that all AoT conditions are identically satisfied. In
particular,
\begin{align}
P(m_i) &= \sum_{m_j=\pm1} P(m_i,m_j), \qquad i<j, \\
P(m_1,m_2) &= \sum_{m_3=\pm1} P(m_1,m_2,m_3),
\end{align}
and their higher-order counterparts hold for arbitrary parameter values.
This reflects the causal structure of the dynamical map.

In contrast, NSIT conditions are generically violated. While certain two-time
NSIT relations,
\begin{equation}
P(m_j)=\sum_{m_i=\pm1}P(m_i,m_j), \qquad i<j,
\end{equation}
may be satisfied in restricted regimes, the three-time NSIT relations,
\begin{align}
P(m_1,m_3) &= \sum_{m_2=\pm1} P(m_1,m_2,m_3), \\
P(m_2,m_3) &= \sum_{m_1=\pm1} P(m_1,m_2,m_3),
\end{align}
are violated in general by the hybrid evolution.

The failure of NSIT, despite universal satisfaction of AoT constraints,
precludes the existence of a global joint probability distribution. The hybrid
Liouvillian is therefore intrinsically incompatible with macroscopic realism.

.





\begin{thebibliography}{99}

\bibitem{Leggett2002} 
A. J. Leggett, J. Phys.: Condens. Matter \textbf{14}, R415 (2002).

\bibitem{Schlosshauer2005} 
M. Schlosshauer, Rev. Mod. Phys. \textbf{76}, 1267 (2005).

\bibitem{Zurek2003} 
W. H. Zurek, Rev. Mod. Phys. \textbf{75}, 715 (2003).

\bibitem{Kofler2007} 
J. Kofler and C. Brukner, Phys. Rev. Lett. \textbf{99}, 180403 (2007).

\bibitem{LG1985} 
A. J. Leggett and A. Garg, Phys. Rev. Lett. \textbf{54}, 857 (1985).

\bibitem{Leggett1988} 
A. J. Leggett, Found. Phys. \textbf{18}, 939 (1988).

\bibitem{Kofler2013}
J. Kofler and \v{C}. Brukner,
Phys. Rev. A \textbf{87}, 052115 (2013).

\bibitem{Hensen2015}
B. Hensen et al.,
Nature \textbf{526}, 682-686 (2015).

\bibitem{Bell1964} 
J. S. Bell, Physics \textbf{1}, 195 (1964).

\bibitem{Brunner2014} 
N. Brunner, D. Cavalcanti, S. Pironio, V. Scarani, and S. Wehner, Rev. Mod. Phys. \textbf{86}, 419 (2014).

\bibitem{Emary2014} 
C. Emary, N. Lambert, and F. Nori, Rep. Prog. Phys. \textbf{77}, 016001 (2014).

\bibitem{Palacios2010} 
A. Palacios-Laloy, A. F. Mallet, F. Nguyen, P. Bertet, D. Vion, D. Esteve, and A. N. Korotkov, Nat. Phys. \textbf{6}, 442 (2010).

\bibitem{Groblacher2015} 
S. Groblacher, A. Trubarov, N. Prigge, G. D. Cole, M. Aspelmeyer, and J. Eisert, Nature (London) \textbf{520}, 531 (2015).

\bibitem{Katiyar2013} 
H. Katiyar, A. Shukla, K. R. K. Rao, and T. S. Mahesh, Phys. Rev. A \textbf{87}, 052102 (2013).

\bibitem{Mouse2017} 
K. R. K. Rao, H. Katiyar, T. S. Mahesh, A. Sen, U. Sen, and A. Kumar, Phys. Rev. A \textbf{95}, 022104 (2017).

\bibitem{Waldherr2011} 
G. Waldherr, P. Neumann, S. F. Huelga, F. Jelezko, and J. Wrachtrup, Phys. Rev. Lett. \textbf{107}, 090401 (2011).

\bibitem{Fritz2010} 
T. Fritz, New J. Phys. \textbf{12}, 083055 (2010).

\bibitem{Budiyono2013} 
A. Budiyono and C. Emary, Phys. Rev. A \textbf{87}, 032103 (2013).

\bibitem{Maroney2014} 
O. J. E. Maroney and C. G. Timpson, arXiv:1412.6139 (2014).


\bibitem{Varma2021}
A. V. Varma, I. Mohanty, and S. Das, J. Phys. A: Math. Theor. 54, 115301 (2021).

\bibitem{KumariPan1}
Javid Naikoo et al 2021 J. Phys. A: Math. Theor. 54 275303


\bibitem{KumariPan2}
A.Kumari, A. K.Pan, ANNALEN DER PHYSIK2022, 534, 2100401. 

\bibitem{Dressel2014} 
J. Dressel, M. Malik, F. M. Miatto, A. N. Jordan, and R. W. Boyd, Rev. Mod. Phys. \textbf{86}, 307 (2014).

\bibitem{Knee2016} 
G. C. Knee, K. Kakuyanagi, M.-C. Yeh, Y. Matsuzaki, H. Toida, H. Yamaguchi, S. Saito, A. J. Leggett, and W. J. Munro, Nat. Commun. \textbf{7}, 13253 (2016).

\bibitem{Jordan2015} 
A. N. Jordan, J. Tollaksen, J. E. Troupe, J. Dressel, and Y. Aharonov, Quantum Stud.: Math. Found. \textbf{2}, 5 (2015).

\bibitem{VarmaYogesh}
A. V. Varma et al., Phys. Rev. A 108, 032202 (2023).

\bibitem{SpaulJPhysA}
S. Paul, A. V. Varma, and S. Das, 
Leggett–Garg inequality, 
J. Phys. A: Math. Theor. 57, 385203 (2024).

\bibitem{SpaulPRAletters}
S. Paul, A. V. Varma, Y. N. Joglekar, and S. Das,
Phys. Rev. A 113, L040603 (2026).

\bibitem{Minganti2019}
F.~Minganti, A.~Miranowicz, R.~W. Chhajlany, and F.~Nori,
Phys. Rev. A \textbf{100}, 062131 (2019).
\bibitem{Kumar2020}
P. Kumar, K. Snizhko, and Y. Gefen, 
Phys. Rev. A \textbf{101}, 062112 (2020).

\bibitem{Kumar2021}
P. Kumar, K. Snizhko, and Y. Gefen, 
Phys. Rev. A \textbf{104}, L050405 (2021).




\bibitem{Gorini1976} 
V. Gorini, A. Kossakowski, and E. C. G. Sudarshan, J. Math. Phys. \textbf{17}, 821 (1976).

\bibitem{Lindblad1976} 
G. Lindblad, Commun. Math. Phys. \textbf{48}, 119 (1976).

\bibitem{Breuer2002} 
H.-P. Breuer and F. Petruccione, \textit{The Theory of Open Quantum Systems} (Oxford University Press, Oxford, 2002).

\bibitem{Plenio1998} 
M. B. Plenio and P. L. Knight, Rev. Mod. Phys. \textbf{70}, 101 (1998).

\bibitem{Carmichael1993} 
H. J. Carmichael, \textit{An Open Systems Approach to Quantum Optics} (Springer-Verlag, Berlin, 1993).

\bibitem{Wiseman2009} 
H. M. Wiseman and G. J. Milburn, \textit{Quantum Measurement and Control} (Cambridge University Press, Cambridge, 2009).

\bibitem{Jacobs2014} 
K. Jacobs, \textit{Quantum Measurement Theory and its Applications} (Cambridge University Press, Cambridge, 2014).


\bibitem{MaheshUshaDevi2024} 
A. Chatterjee, G.S. Karthik, T.S. Mahesh, A. R. Usha Devi, Phys. Rev. Lett. \textbf{135}, 220202 (2025).

\bibitem{AKPan}
A. K. Pan, Phys. Rev. A 102, 032206 (2020).

\bibitem{Majidy}
S.-S. Majidy, H. Katiyar, G. Anikeeva, J. Halliwell, and R.
Laflamme, Phys. Rev. A 100, 042325 (2019).

\bibitem{Huffman}
E. Huffman and A. Mizel, Phys. Rev. A 95, 032131 (2017).

\bibitem{Clemente}
L. Clemente and J. Kofler, Phys. Rev. A 91, 062103 (2015).

\bibitem{Luders1951} 
G. Lüders, Ann. Phys. (Leipzig) \textbf{443}, 322 (1951).

\bibitem{Budroni2014} 
C. Budroni and C. Emary, Phys. Rev. Lett. \textbf{113}, 050401 (2014).

\bibitem{Rotter2009} 
I. Rotter, J. Phys. A: Math. Theor. \textbf{42}, 153001 (2009).

\bibitem{Daley2014} 
A. J. Daley, Adv. Phys. \textbf{63}, 77 (2014).

\bibitem{Childs2012} 
A. M. Childs and N. Wiebe, Quantum Info. Comput. \textbf{12}, 901 (2012).

\bibitem{Berry2015} 
D. W. Berry, A. M. Childs, R. Cleve, R. Kothari, and R. D. Somma, Phys. Rev. Lett. \textbf{114}, 090502 (2015).

\bibitem{Bender2007} 
C. M. Bender, Rep. Prog. Phys. \textbf{70}, 947 (2007).

\bibitem{Tang2018} 
J.-S. Tang, Y.-T. Wang, S. Yu, D.-Y. He, J.-S. Xu, B.-H. Liu, G. Chen, Y.-N. Sun, K. Sun, Y.-J. Han, et al., Nat. Photon. \textbf{10}, 642 (2016).

\bibitem{Aharonov1988} 
Y. Aharonov, D. Z. Albert, and L. Vaidman, Phys. Rev. Lett. \textbf{60}, 1351 (1988)

\bibitem{Gilyen2019} 
A. Gilyén, Y. Su, G. H. Low, and N. Wiebe, in \textit{Proceedings of the 51st Annual ACM SIGACT Symposium on Theory of Computing} (ACM, New York, 2019), p. 193.
\end{thebibliography}
\end{document}